\documentclass[pdftex,twocolumn,epjc3]{svjour3}        
\def\makeheadbox{}
\usepackage{here}	
\usepackage{graphicx}				
\usepackage{amsmath,amssymb,bm}			
\usepackage{caption}
\usepackage{graphics}
\usepackage{soul}
\usepackage{url}
\usepackage{float}
\usepackage{multirow}
\usepackage{array,booktabs}	
\usepackage{lineno}
\linenumbers
\usepackage{physics}
\usepackage{placeins}
\usepackage{float}
\usepackage[Export]{adjustbox}
\usepackage{pifont}
\usepackage{comment}
\usepackage{rotate}
\usepackage[overlay]{textpos}
\usepackage{braket}
\usepackage{tabularx}
\usepackage[utf8]{inputenc}

\usepackage{times}
\usepackage{tipa}
\usepackage{xspace}

\usepackage{color}
\usepackage[normalem]{ulem}
\usepackage{cancel}
\usepackage{txfonts}

\usepackage{float} %
\usepackage[caption=false]{subfig} %

\usepackage{tablefootnote}

\usepackage{scalerel}
\usepackage{tikz}
\usetikzlibrary{svg.path}

\definecolor{orcidlogocol}{HTML}{A6CE39}
\tikzset{
  orcidlogo/.pic={
    \fill[orcidlogocol] svg{M256,128c0,70.7-57.3,128-128,128C57.3,256,0,198.7,0,128C0,57.3,57.3,0,128,0C198.7,0,256,57.3,256,128z};
    \fill[white] svg{M86.3,186.2H70.9V79.1h15.4v48.4V186.2z}
                 svg{M108.9,79.1h41.6c39.6,0,57,28.3,57,53.6c0,27.5-21.5,53.6-56.8,53.6h-41.8V79.1z M124.3,172.4h24.5c34.9,0,42.9-26.5,42.9-39.7c0-21.5-13.7-39.7-43.7-39.7h-23.7V172.4z}
                 svg{M88.7,56.8c0,5.5-4.5,10.1-10.1,10.1c-5.6,0-10.1-4.6-10.1-10.1c0-5.6,4.5-10.1,10.1-10.1C84.2,46.7,88.7,51.3,88.7,56.8z};
  }
}

\newcommand\orcidicon[1]{\href{https://orcid.org/#1}{\mbox{\scalerel*{
\begin{tikzpicture}[yscale=-1,transform shape]
\pic{orcidlogo};
\end{tikzpicture}
}{|}}}}

\usepackage{cite}

\usepackage[colorlinks=true, citecolor=blue, linkcolor=blue, urlcolor=blue, linktoc=all]{hyperref}

\author{
Allencris~John~Rubesh~Rajan\thanksref{addr1,addr2} \protect\orcidicon{0000-0002-5397-6782}
        \and
        Laboni~Manna\thanksref{addr3}\protect\orcidicon{0000-0002-8559-5108}
}

\institute{Universit\'e Paris-Saclay, CNRS, IJCLab, 91405 Orsay, France \label{addr1}
\and
School of Physics, University College Dublin,
Dublin 4, Ireland \label{addr2}
           \and
           Warsaw University of Technology, plac Politechniki 1, Warsaw, Poland \label{addr3}
           \\~ \\
\email{laboni.manna.dokt@pw.edu.pl \& allen.johnrubeshrajan@ucdconnect.ie}
}

\date{\today}
\nolinenumbers
\begin{document}

\title{Prospects for early heavy-quark measurements at the EIC}

\maketitle
\begin{abstract}
We assess the prospects for inclusive heavy-quark-pair electroproduction in electron-proton ($ep$) and electron-nucleus ($e$A) collisions, alongside double-charm-pair ($D^0 \bar{D}^0$ and $D^0D^0$) electroproduction in $ep$ collisions, during the early operational phase of the Electron-Ion Collider (EIC). Based on the anticipated beam energies and luminosities of the EIC, we estimate the total cross sections and differential distributions for photon virtuality $Q^2$, heavy-quark transverse momentum, and rapidity. This analysis evaluates different Bjorken-$x$ ($x_{\rm Bj}$) intervals across a range of $Q^2$ values. Expected event yields are estimated from the projected integrated luminosities of the early science programme, demonstrating the substantial improvement in statistical precision over previous HERA measurements at moderate and large $x_{{\rm Bj}}$. For electron--nucleus collisions, we compute the nuclear modification ratio $R^{e\text{A}}$ within collinear factorisation using nuclear parton distribution functions, and identify the kinematic regimes in which shadowing, anti-shadowing, and EMC-like suppression are accessible with early EIC data. These results are intended to serve as a parton-level baseline to support feasibility studies and to provide a reference for future higher-order theoretical and experimental investigations at the EIC.
\end{abstract}

\tableofcontents

\section{Introduction}\label{sec:intro}

Inclusive electroproduction of heavy-flavour (HF) in deep inelastic scattering (DIS) off nucleons and nuclei plays a central role in establishing the partonic structure of hadrons and in testing perturbative quantum chromodynamics (pQCD) over a wide range of energy scales. In particular, the inclusive electroproduction of heavy quarks provides a clean and theoretically well-controlled probe of QCD dynamics. While photon virtuality $Q^2$ sets the hard scale in inclusive DIS, the heavy-quark mass $m_{Q}$ introduces an additional intrinsic hard scale, ensuring the applicability of pQCD even in the low-$Q^2$ regime. 
Moreover, while light-hadron DIS accesses gluons only indirectly through DGLAP evolution, HF production in DIS proceeds predominantly via virtual photon--gluon ($\gamma^* g \to Q \bar{Q}$) fusion at leading order (LO) in a fixed-flavour number scheme (FFNS)~\cite{Forte:2010, Harris:1999}, providing direct---though scheme-dependent---sensitivity to the gluon content of the target in both proton and nuclear environments. While our calculations are performed within FFNS, variable flavour number schemes (VFNS) offer an equally well-established complementary framework, with recent work further refining heavy-quark mass effects in DIS structure functions~\cite{Ball:2024vfns,Hekhorn:2019nlf, Hekhorn:2024tqm}.

Within DIS processes, inclusive HF electroproduction occupies a distinctive position due to the presence of multiple hard scales. At moderate and large photon virtualities, the heavy-quark mass and the virtuality $Q^2$ jointly determine the perturbative expansion, allowing for quantitative tests of fixed-order calculations and mass-dependent schemes. 
At lower values of $Q^2$ and $x_{\mathrm{Bj}}$, threshold mass effects compete directly with small-$x_{\mathrm{Bj}}$ dynamics, making cross-section measurements a sensitive probe of both the target gluon distribution and its low-$x_{\mathrm{Bj}}$ evolution~\cite{Laenen:1992zk}.
 
As a result, charm and beauty production in DIS provide important constraints on parton distribution functions and serve as benchmark observables for validating QCD factorisation in regions where multiple scales are relevant.

At the HERA collider, the H1 and ZEUS experiments performed extensive studies of inclusive HF heavy-quark electroproduction in $ep$ collisions~\cite{H1:2009uwa,H1:2018flt,H1:2012xnw,H1:2015ubc,Schmitt:2017nwe,H1:2009pze,H1:2005vma,H1:2004esl,ZEUS:2007yva,H1:2004bwe,ZEUS:1997tfb}. 
Inclusive and semi-inclusive measurements of charm and beauty production were carried out over the range $2.5 \leq Q^2 \leq 400~\mathrm{GeV}^2$ and $3\times10^{-5} \lesssim x_\text{Bj} \lesssim 5\times10^{-2}$~\cite{H1:2018flt}, yielding important constraints on the proton's gluon distribution thereby providing essential input to global PDF analyses~\cite{Hou:2019efy,H1:2015ubc}. 
These measurements provided a stringent test of perturbative QCD calculations for heavy-quark structure functions, confronting precision collider data with fixed-order calculations~\cite{Laenen:1992zk,Buza:1995ie} and their subsequent refinements, up to complete $\mathcal{O}(\alpha_{s}^{3})$ corrections to the HF structure functions~\cite{Ablinger:2014vwa, Ablinger:2025awb}. 

While these milestones established the field, HF measurements at HERA still lacked statistical precision, particularly at low $x_{{\rm Bj}}$ and low-to-moderate $Q^2$~\cite{Abramowicz:2015mha,H1:2018flt}, as well as for multi-differential distributions in heavy-quark ($Q$) transverse momentum, $p_T^{Q}$, and heavy-quark rapidity, $y_{Q}$. In this kinematic regime, lower production cross sections, decreased tagging efficiencies, and restricted detector acceptance led to larger uncertainties than in the inclusive case~\cite{Aaron:2009aa,ZEUS:2014gka}, reducing the discriminating power of the HF data in global PDF fits~\cite{Bailey:2020ooq,Hou:2019efy} relative to what would be achievable with higher luminosity.
In particular, the relative statistical uncertainty on the combined charm reduced cross section reached the level of 10--20\% at low $x_{{\rm Bj}}$ and low $Q^2$, and degrades significantly for multi-differential distributions in $p_T^{Q}$ and $y_{Q}$~\cite{H1:2018flt}.
In addition, while a broad programme of electron- and 
muon-nucleus DIS experiments was carried out at fixed-target 
facilities such as EMC~\cite{EuropeanMuon:1983wih}, NMC~\cite{NewMuon:1993oys}, SMC~\cite{SpinMuonSMC:1997mkb} and HERMES~\cite{HERMES:1998mat}, these 
measurements focused on inclusive and semi-inclusive structure 
functions of light quarks.  
Inclusive heavy-quark pair production in DIS on nuclear targets was not measured in any of these fixed-target programmes (nor at HERA, which operated solely with proton beams), leaving the nuclear gluon distribution very poorly constrained~\cite{AbdulKhalek:2021gbh}. Complementary opportunities to study HF production and nuclear DIS at lower centre-of-mass energies are offered by the proposed Electron-Ion Collider in China (EicC)~\cite{Anderle:2021eicc,Anderle:2023uvi}.

The future EIC is designed to substantially extend the DIS HF programme by combining high luminosity with a wide range of centre-of-mass energies and the capability to collide electrons with both protons and nuclei. The projected luminosities exceed those achieved at HERA by one to two orders of magnitude, enabling precision measurements of heavy-quark production in kinematic regions that were previously statistics-limited~\cite{AbdulKhalek:2021gbh}. In particular, the EIC will allow for systematic studies of HF electroproduction across a broad range of $x_{{\rm Bj}}$ and $Q^2$, including regions relevant for the transition between moderate and small $x_{{\rm Bj}}$, as well as detailed investigations of transverse-momentum and rapidity distributions. In addition, HF electroproduction has been studied as a probe of gluon helicity at the EIC~\cite{Anderle:2021hpa,Hekhorn:2024tqm}, where the unpolarised cross sections presented here provide the necessary baseline for future asymmetry measurements.

Heavy-quark pair production is a cornerstone of the EIC nuclear physics programme~\cite{AbdulKhalek:2021gbh}. In $e$A collisions, charm and beauty electroproduction offer a theoretically clean probe 
of nuclear parton distribution functions (nPDFs)~\cite{Eskola:2009uj,Eskola:2021nhw,Klasen:2023,Kovarik:2015cma,AbdulKhalek:2022fyi}, benefiting from reduced sensitivity to hadronisation effects relative to light-flavour measurements. Crucially, as highlighted by nPDF reweighting and global fit analyses using quarkonium, beauty and charm production data in proton-nucleus collisions at RHIC and LHC ~\cite{Kusina:2017gkz,Kusina:2020dki,Duwentaster:2021icx,AbdulKhalek:2022fyi}, HF probes carry remarkable sensitivity to the nuclear gluon distribution. 
EIC HF measurements will not only constrain small-$x_{{\rm Bj}}$ gluon shadowing, but will also pin down the elusive gluon anti-shadowing region and shed light on potential gluon EMC effects at larger $x_{{\rm Bj}}$~\cite{Malace:2014uea,Klasen:2023ugq,Segarra:2020gtj,Cloet:2019mql,Arrington:2021vuu}.
Establishing a precise baseline for the gluon density in nuclei is indispensable for disentangling initial-state nPDF modifications from other cold nuclear matter effects~\cite{Gerschel:1988wn,Vogt:1999cu,Ferreiro:2014bia,Capella:2005cn,Capella:2000zp,Gavin:1990gm,Arleo:2012hn,Sharma:2012dy,Arleo:2010rb,Brodsky:1992nq,Gavin:1991qk,Brodsky:1989ex,Ducloue:2015gfa,Ma:2015sia,Fujii:2013gxa,Qiu:2013qka,Kopeliovich:2001ee,Ferreiro:2008wc,Ferreiro:2011xy,Ferreiro:2013pua,Vogt:2010aa} on hard probes in order to probe quark-gluon plasma with heavy quarks and quarkonia in nucleus-nucleus collisions at RHIC and the LHC~\cite{Andronic:2015wma,Chapon:2020heu}. 

To this end, 
evaluating $e$A electroproduction observables within the collinear factorisation framework provides such estimates of the baseline nuclear modifications of the gluon density via the ratio $R^{e\text{A}}$. While non-linear effects such as small-$x_\text{Bj}$ saturation become relevant at the lowest $x_\text{Bj}$ accessed at the EIC, the present results provide a collinear factorisation baseline against which future higher-order calculations and EIC measurements can be compared against, and thanks to which one can better assess the accessible kinematic range (in $Q^2$,   $x_\text{Bj}$, $y_{Q}$, $p_T^{Q}$ etc.) where HF probes can be measured at the EIC .

This work provides theoretical predictions for inclusive charm and beauty electroproduction ($c\bar{c}$, $b\bar{b}$) in $ep$ and $e$A collisions using copper (Cu), silver (Ag), and gold (Au) nuclear targets for early EIC operations. In addition, we present theoretical predictions for inclusive double-charm-pair electroproduction. By evaluating total cross sections and $x_\text{Bj}$ ranges alongside differential distributions in photon virtuality $Q^2$, heavy-quark transverse momentum $p_T^{Q}$, and heavy-quark rapidity $y_{Q}$, we quantify the accessible kinematic reach and expected statistical precision achievable with early-science luminosities. 

This paper is organised as follows: in Section~\ref{sec:dis}, we recall DIS kinematics to define the key variables and observables, followed by the computational setup in Section~\ref{sec:setup}. In Section~\ref{sec:results}, we present results for $ep$, $e$Cu, $e$Ag and $e$Au collisions and extend our analysis to double-charm-pair ($D^0 \bar{D}^0$ and $D^0 D^0$) electroproduction in $ep$ collisions. Finally, we present our conclusions and outlook in Section~\ref{sec:outlook}.

\section{DIS kinematics and observables}\label{sec:dis}

\begin{figure}[t]
    \centering
    \includegraphics[height=6.5cm, keepaspectratio]
    {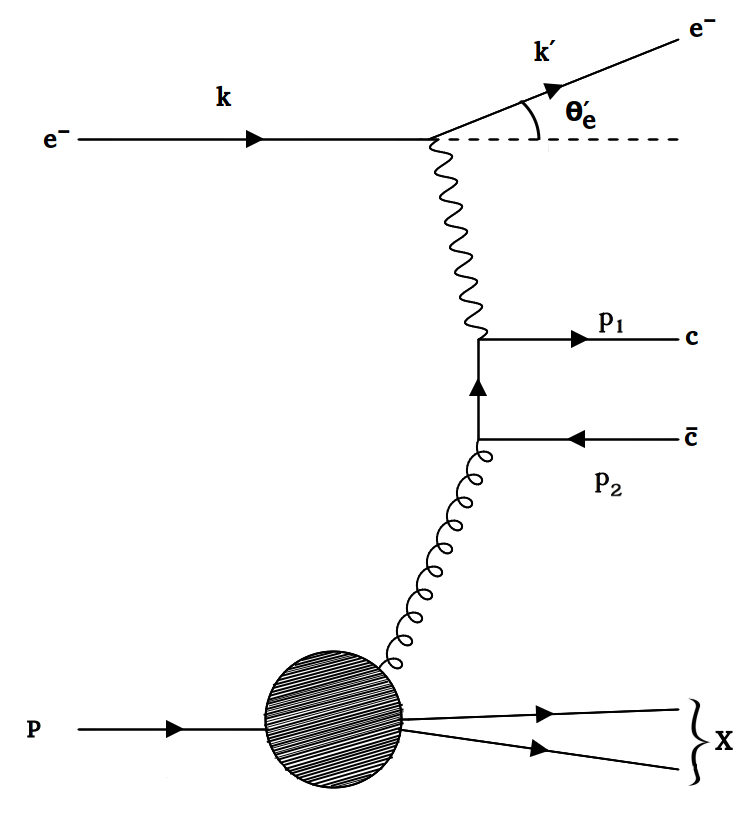}
    \caption{Representative Feynman diagram illustrating the inclusive DIS process $e + p \to e + c\bar{c} + X$ at Born order in a 3-flavour scheme, along with the four-momentum assignments adopted in this work.}
    \label{fig:dis_kinematics}
\end{figure}

We consider inclusive deep inelastic scattering of electrons off 
protons or nuclei,
\begin{equation}
    e(k) + p(P) \to e(k^{\prime}) + {Q}(p_1) + 
\bar{{Q}}(p_2) + X,
\end{equation}
where $k$ and $k'$ denote the four-momenta of the incoming and scattered electrons, $P$ is the four-momentum of the incoming  hadron, and $p_i$ represents the four-momenta of the heavy-quark pair. The exchanged virtual photon carries four-momentum $q = k - k^{\prime}$, and we define the standard DIS variable $Q^2 \equiv -q^2 > 0$. The process and its momentum assignments are schematically depicted via the Feynman diagram in 
Fig.~\ref{fig:dis_kinematics}.

We work in the laboratory frame where the incoming electron has 
energy $E_e$, the scattered electron has energy $E^{\prime}_{e}$, 
and $\theta^{\prime}_{e}$ denotes the electron scattering angle 
with respect to the incoming electron beam direction. The standard 
DIS kinematic variables are given by~\cite{Devenish:2004pb}
\begin{equation}
Q^{2} = 4 E_e E'_e \sin^{2}\left(\frac{\theta'_e}{2}\right),
\end{equation}
which sets the hard scale of the process, and
\begin{equation}
y = \frac{P \cdot q}{P \cdot k}
= 1 - \frac{E'_e}{E_e}\cos^{2}\left(\frac{\theta'_e}{2}\right),
\end{equation}
where $y$ is the inelasticity, corresponding to the fractional 
energy loss of the electron in the hadron rest frame. The Bjorken 
scaling variable $x_{\rm Bj}$ is given by
\begin{equation}
x_{\rm Bj} = \frac{Q^{2}}{2\,P\cdot q}
= \frac{Q^{2}}{y\,s_{ep}},
\end{equation}
with $s_{ep} = (k + P)^2$ the squared electron--hadron centre-of-mass (c.m.) energy. In the parton model, $x_{\rm Bj}$ can be interpreted as the longitudinal momentum fraction of the struck parton inside the hadron at LO~\cite{Bjorken:1968dy}. The invariant mass of the photon-hadron system,
\begin{equation}
    W^2 = (P+q)^2 = Q^{2} \left( \frac{1-x_{\rm Bj}}
    {x_{\rm Bj}} \right),
\end{equation}
determines the available phase space for heavy-quark pair production; a minimum $W > 2m_{Q}$ is required to produce a ${Q}\bar{{Q}}$ pair.

For final-state particles, we characterise the kinematics using 
the rapidity
\begin{equation}
    y_{{Q}_i} = \frac{1}{2}\ln\frac{E_i + p_{z,i}}{E_i - p_{z,i}},
\end{equation}
and the pseudorapidity
\begin{equation}
    \eta_{{Q}_i} = -\ln\left(\tan\frac{\theta_i}{2}\right),
\end{equation}
where $E_i$, $p_{z,i}$, and $\theta_i$ are the energy, longitudinal momentum component, and polar angle of particle $i$ with respect to the hadron beam direction, which we take to be the positive $z$ direction. The transverse momentum of each heavy quark is defined as
\begin{equation}
    p^{Q}_{T,i} = \sqrt{p_{x,i}^2 + p_{y,i}^2},
\end{equation}
where $p_{x,i}$ and $p_{y,i}$ are the momentum components transverse to the beam axis. For a single heavy quark ${Q}$ we denote its rapidity $y_{Q}$, while $y_{{Q}\bar{{Q}}}$ and $p_{T,{Q}\bar{{Q}}}$ denote the rapidity and transverse momentum of the reconstructed heavy-quark pair. While rapidity and pseudorapidity ($\eta_{Q}$) coincide in the massless limit, they differ appreciably for massive heavy quarks at moderate $p_{T}^{Q}$. We therefore use rapidity $y_Q$ throughout the results presented in Section~\ref{sec:results}.

The partonic cross sections computed in this work are directly related to the charm and beauty contributions to the proton structure function $F_2$. In the one-photon exchange approximation, the double-differential cross section for heavy-quark electroproduction is given by~\cite{Devenish:2004pb}
\begin{equation}
\begin{aligned}
\frac{d^2 \sigma^{{Q}\bar{{Q}}}}{d x_{\text{Bj}} \, d Q^2} = \frac{2\pi \alpha_{\text{em}}^2}{Q^4 x_{\text{Bj}}} \Big[ &\left(1 + (1 - y)^2\right) F_2^{{Q}\bar{{Q}}}(x_{\text{Bj}}, Q^2) \\
&- y^2 F_L^{{Q}\bar{{Q}}}(x_{\text{Bj}}, Q^2) \Big]
\end{aligned}
\end{equation}
where $\alpha_\text{em}$ is the electromagnetic coupling constant and $F_L^{{Q}\bar{{Q}}}$ is the longitudinal structure function. The contribution from $F_L^{{Q}\bar{{Q}}}$ is suppressed at low $y$ and is typically small compared to $F_2^{{Q}\bar{{Q}}}$ across the kinematic range studied here~\cite{H1:2018flt}. The structure functions $F_2^{c\bar{c}}$ and $F_2^{b\bar{b}}$, which were the primary observables extracted from HF measurements at HERA~\cite{H1:2009uwa,H1:2018flt}, are therefore directly accessible from the differential distributions presented in Section~\ref{sec:results}. Since heavy-quark production in DIS proceeds at leading order via photon-gluon fusion\footnote{In a scheme where the considered heavy quark is not present in the proton.}, which is precisely the subprocess depicted in Fig.~\ref{fig:dis_kinematics}, these structure functions provide a direct probe of the gluon distribution in the target, making $F_2^{c\bar{c}}$ and $F_2^{b\bar{b}}$ among the most theoretically clean gluon-sensitive observables accessible at the EIC.

To evaluate cold nuclear matter effects in $e$A collisions, we construct the per-nucleon nuclear modification ratio:
\begin{equation}
R^{eA} = \frac{1}{A} \frac{\sigma^{eA}}{\sigma^{ep}}
\label{r_eA}
\end{equation}
Here, $R^{eA}$ is defined specifically for heavy-quark production. Within the FFNS adopted in this work, deviations of $R^{eA}$ from unity at LO directly reflect modifications to the nuclear gluon distribution probed via virtual photon–gluon fusion\footnote{At higher orders or in other schemes, such as the variable-flavour-number scheme (VFNS), initial-state heavy quarks and additional subchannels also contribute.}. 
As a function of  $x_{\rm Bj}$, $R^{e\text{A}}$ spans several distinct physical regimes: a depletion at small $x_{\rm Bj}$ ($x_{\rm Bj} \lesssim 0.05$) due to shadowing, followed by an anti-shadowing enhancement in the range $0.05 \lesssim x_{{\rm Bj}} \lesssim 0.3$. At intermediate values ($0.3 \lesssim x_{{\rm Bj}} \lesssim 0.7$), bound-nucleon modifications yield the characteristic suppression of the EMC effect~\cite{EuropeanMuon:1983wih,Malace:2014uea}, whereas intrinsic Fermi motion causes a rapid rise at large $x_{{\rm Bj}}$ ($x_{{\rm Bj}} \gtrsim 0.7$).

\section{Computational Setup}\label{sec:setup}

To compute the heavy-quark production cross sections in DIS, we employ \texttt{MadGraph5\_aMC@NLO}~\cite{Alwall:2014hca}, which provides an automated framework for generating fixed-order matrix elements within collinear factorisation. Currently, DIS processes in \texttt{MadGraph5\_aMC@NLO} are limited to LO accuracy. Consequently, we focus on establishing baseline cross sections and kinematic distributions to serve as a benchmark for the early EIC physics programme.
\begin{table*}[h!]
\centering
\caption{EIC Early Science Matrix showing the 
planned beam species, energies, integrated luminosities, and 
polarisation configurations for the first five years of 
operation~\cite{Arleo:2026tpb}. The $eA$ luminosity is 
given per nucleon.
}
\label{tab:early-science-matrix}
\begin{tabular}{c l c c c c}
\toprule
\textbf{Year} 
& \textbf{Collision system} 
& \textbf{Beam energies (GeV)} 
& \textbf{Luminosity (fb$^{-1}$/year)} 
& \textbf{$e^-$ polarisation} 
& \textbf{$p/A$ polarisation} \\
\midrule
\textbf{1} 
& $e$Ag, Ru or Cu 
& $10 \times 115$ 
& $0.9$ 
& NO (commissioning)
& N/A \\
\midrule
\multirow{2}{*}{\textbf{2}} 
& $e$D 
& $10 \times 130$ 
& $13.1$ 
& \multirow{2}{*}{LONG} 
& NO \\
& $ep$
& $10 \times 130$ 
& $5.6$--$6.1$ 
& 
& TRANS \\
\midrule
\textbf{3} 
& $ep$
& $10 \times 130$ 
& $5.6$--$6.1$ 
& LONG 
& TRANS and/or LONG \\
\midrule
\multirow{2}{*}{\textbf{4}} 
& $e$Au 
& $10 \times 100$ 
& $0.95$ 
& \multirow{2}{*}{LONG} 
& N/A \\
& $ep$
& $10 \times 250$ 
& $7.1$--$10.6$ 
& 
& TRANS and/or LONG \\
\midrule
\multirow{2}{*}{\textbf{5}} 
& $e$Au 
& $10 \times 100$ 
& $0.95$ 
& \multirow{2}{*}{LONG} 
& N/A \\
& $e{}^3$He 
& $10 \times 166$ 
& $9.8$ 
& 
& TRANS and/or LONG \\
\bottomrule
\end{tabular}
\vspace{-0.5 cm}
\end{table*}
We adopt the three-flavour number scheme (3FNS) for charm production and the four-flavour number scheme (4FNS) for beauty production, treating the heavy quarks as massive final-state particles throughout. Alternative treatments within the FONLL framework have also been applied to HF production at the EIC~\cite{Hekhorn:2024tqm}. The renormalisation and factorisation scales are set to a common central value $\mu_0 = H_T/2$, where $H_T$ is the scalar sum of the transverse masses of the final-state heavy quarks. 
To estimate the dominant theoretical uncertainty associated with the scale choice, we perform a standard nine-point variation by independently varying $\mu_R$ and $\mu_F$ by factors of two around $\mu_0$, i.e., $\mu_{R,F} \text{ in } \{0.5,1,2\} \mu_0$.Parton distribution functions are taken from the CT18ANLO set~\cite{Hou:2019efy}. For electron–nucleus collisions, while our study primarily uses the EPPS21 nuclear PDF set~\cite{Eskola:2021nhw}, the analysis is equally applicable to other modern nuclear PDF sets, such as nCTEQ~\cite{Muzakka:2022has,Kovarik:2015cma} and nNNPDF~\cite{AbdulKhalek:2022fyi,AbdulKhalek:2019mzd,AbdulKhalek:2020yuc}. The charm- and beauty-quark masses are set to $m_c = 1.55~\mathrm{GeV}$ and $m_b = 4.75~\mathrm{GeV}$, respectively~\cite{ParticleDataGroup:2024cfk,Alekhin:2012vu}.

To enable direct comparison with existing and future experimental data, we apply parton-level fiducial cuts designed to emulate the acceptance of the proposed ePIC detector~\cite{Li:2025lxr}. We require the scattered electron pseudorapidity to satisfy $\vert{}\eta_e\vert{} < 3.5$, and an acceptance cut of $\vert{}\eta_Q\vert{} < 3.5$ is applied to the final-state heavy quarks, consistent with the projected ePIC detector coverage for HF reconstruction~\cite{Li:2025lxr}.

The EIC will operate with a staged physics programme during its early 
years, featuring multiple beam-energy configurations and collision 
systems~\cite{Arleo:2026tpb}. In this study, we focus on a 
representative subset of the planned early running scenarios, 
summarised in Table~\ref{tab:early-science-matrix}. In particular, we 
consider $ep$ and $e$A collisions with different ions at 
c.m.~energies relevant for the early science programme, and 
use the corresponding projected integrated luminosities to estimate 
expected event yields. Unless otherwise stated, the results presented 
below correspond to the $10\times130~\mathrm{GeV}$ $ep$ configuration 
and the $10\times100(115)~\mathrm{GeV}$ $e$A configuration, 
giving c.m.~energies of $\sqrt{s_{ep}} = 72~\mathrm{GeV}$ 
and $\sqrt{s_{eN}} = 63, 67.8~\mathrm{GeV}$ respectively.

To estimate the kinematic reach, we define observability lines corresponding to the differential cross-section threshold required to detect $N_{\rm ev}$ events per bin at the projected 
integrated luminosity:\begin{equation}N_\text{ev} = \frac{\mathrm{d}\sigma}{\mathrm{d}X} \times f \times \epsilon \times \Delta X \times \mathcal{L},\end{equation}where $\mathrm{d}\sigma/\mathrm{d}X$ is the differential cross section with respect to a kinematic variable $X$, $\Delta X$ is the bin width, $\mathcal{L}$ is the integrated luminosity, $f$ is the heavy-quark fragmentation fraction, and $\epsilon$ is the detection efficiency. For charm production, we analyse $D^0$ mesons assuming $f(c \to D^0) = 0.542 \pm 0.024$~\cite{Gladilin:2014tba} and $\epsilon = 3\%$, inspired by ECCE performance studies~\cite{Li:2022ewa}. For beauty production, we target $B^0$ mesons using $f(b \to B^0) = (40.4 \pm 0.6)\%$~\cite{ParticleDataGroup:2024cfk} and $\epsilon = 5\%$, following CMS estimates~\cite{CMS:2011oft,CMS:2011pdu}. Since detailed ePIC performance estimates are not yet available, these parameters provide realistic baseline values. The observability limit is set at $N_{\rm ev} = 20$ (labelled as ``20 ev'' in the plots) for a given luminosity\footnote{For $ep$ collisions, we use the average of the projected $5.6\text{--}6.1~\text{fb}^{-1}$ range given in Table~\ref{tab:early-science-matrix}.} and bin width which roughly corresponds to the statistical accuracy of the typical last bins of HERA measurements.

\section{Results}\label{sec:results}

\subsection{$ep$ results}

 We generated one million events for each of the charm and beauty DIS processes, and present below the resulting cross section distributions alongside observability estimates for the projected early-science luminosities.
\begin{figure}[htbp!]
    \centering
    \includegraphics[width=0.82\columnwidth, keepaspectratio]
    {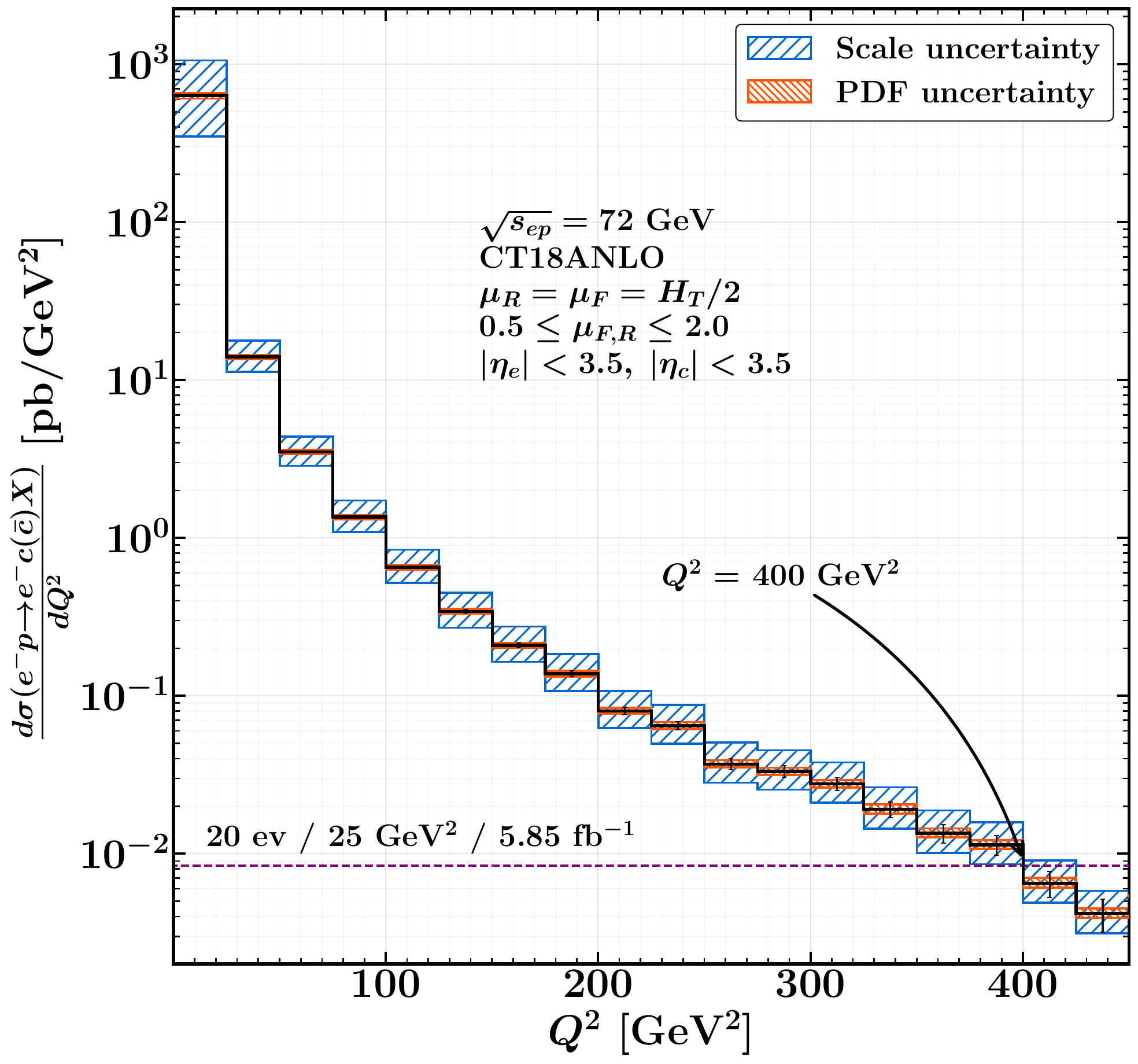}
    \caption{Differential cross section $  d\sigma/dQ^2  $ for the production of a charm quark 
      in $  ep \to e\, c\bar{c} + X$ at a centre-of-mass energy $\sqrt{s_{ep}} = 72$ GeV. The hatched blue and orange band represents the scale and PDF uncertainties respectively, where the scale uncertainty is obtained by varying the factorisation and renormalisation scales in the range $0.5 \leq \mu_{F,R} \leq 2.0$. The purple line indicates the observability line for an integrated luminosity of 5.85 fb$^{-1}$.}
    \label{fig:ep_dsig_dq2_charm}\vspace*{-0.5cm}
\end{figure}

\begin{figure}[htbp!]
    \centering
    \includegraphics[width=0.865\columnwidth, keepaspectratio]
    {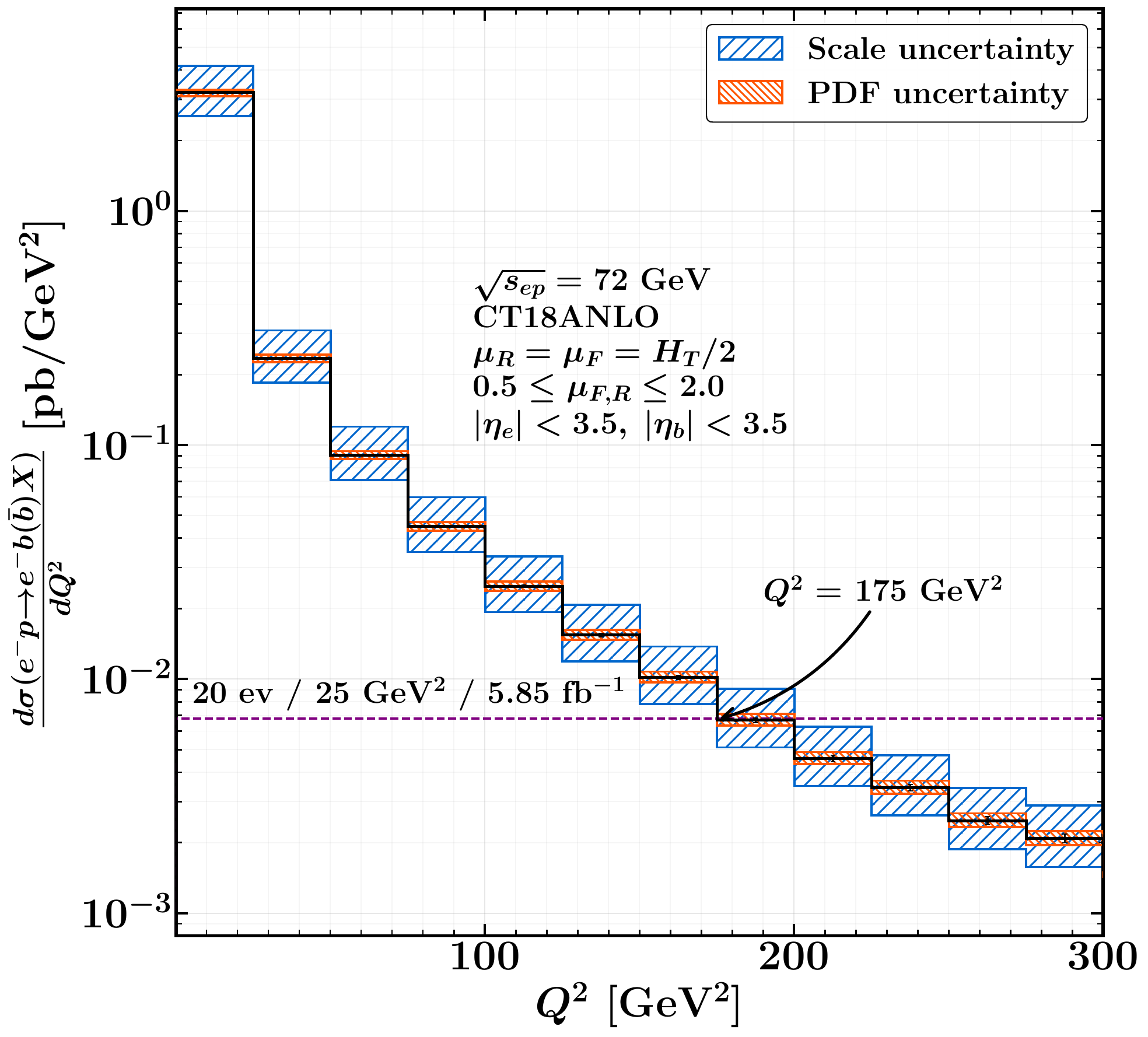} 
    \caption{The differential cross section $  \mathrm{d}\sigma/\mathrm{d}Q^2$ for the production of a beauty quark in $ep \to e\, b\bar{b} + X$ collision at a centre-of-mass energy $\sqrt{s_{ep}} = 72$~GeV. The hatched blue and orange band represents the scale and PDF uncertainties respectively, where the scale uncertainty is obtained by varying the factorisation and renormalisation scales in the range $0.5 \leq \mu_{F,R} \leq 2.0$. The purple line shows the observability line  corresponding to an integrated luminosity of 5.85 fb$^{-1}$.}
    \label{fig:ep_dsig_dq2_bottom} \vspace*{-0.5cm}
\end{figure}
Fig.~\ref{fig:ep_dsig_dq2_charm} shows the ${\rm d}\sigma/{\rm d}Q^2$ distribution for charm production; the corresponding total cross sections and $x_{{\rm Bj}}$ ranges for each $Q^2$ bin are detailed in Table~\ref{tab:heavy_quark_cross_sections}. 
The distribution peaks in the lowest $Q^2$ bin and falls steeply, with the rate of decrease becoming more gradual beyond $Q^2 \approx 100~\text{GeV}^2$ before dropping below the observability threshold near $Q^2 \approx 400~\text{GeV}^2$, a kinematic limit corresponding to $\langle x_{\mathrm{Bj}} \rangle \approx 0.74$\footnote{For each observability threshold across the $Q^2$ and $p_T$ distribution cases, $x_{\mathrm{Bj}}$ is evaluated in the highest kinematic bin. Due to its continuous distribution across this region, we report the weighted average value. Note that, this weighted average value  falls within the $x_{\mathrm{Bj}}$ range reported in Table~\ref{tab:heavy_quark_cross_sections} for the highest observed $Q^2$ interval of 200-400 GeV$^{2}$.}. The corresponding distribution for beauty quark production is presented in Fig.~\ref{fig:ep_dsig_dq2_bottom}. As expected, the cross section for beauty production is significantly smaller than that of charm throughout, driven by the higher beauty quark mass, the associated threshold suppression and the smaller electric charge of the beauty quark relative to the charm.
As shown in Fig.~\ref{fig:ep_dsig_dq2_bottom}, $Q^2$ value up to $175~\text{GeV}^2$ is accessible, which probes $\langle x_{\mathrm{Bj}} \rangle \approx 0.41  $\footnote{Similar to the charm case, this weighted average also falls within the highest $Q^2$ range ($120\text{--}175\text{ GeV}^2$) reported in Table~\ref{tab:heavy_quark_cross_sections}.}. In the high-$Q^2$ regime ($Q^2 \gg m_Q^2$), where mass effects become negligible, the ratio of the beauty to charm cross section approaches a constant value determined by the ratio of their squared electric charges.However, the maximum $Q^2$ values attainable within the present kinematic setup remain insufficient to fully enter this asymptotic regime.
The total cross sections—$\sigma_{c\bar{c}}$  and $\sigma_{b\bar{b}}$—alongside their respective $x_{{\rm Bj}}$ ranges for each $Q^2$ bin, are summarized in Table~\ref{tab:heavy_quark_cross_sections}. For both flavours, the cross section decreases monotonically with increasing $Q^2$, which systematically shifts the accessible region towards larger $x_{{\rm Bj}}$. At low $Q^2$ ($2.5\text{--}5~\mathrm{GeV}^2$), the reach extends down to $x_{{\rm Bj}} \approx 0.0012$ for charm and $0.0010$ for beauty. Overall, $c\bar{c}$ production spans a much broader domain, reaching up to $x_{{\rm Bj}} \approx 0.98$ at high $Q^2$, whereas $b\bar{b}$ production acts as a more localised probe of the small- to intermediate-$x_{{\rm Bj}}$ regime.  Consequently, charm production is better suited for constraining the gluon distribution in global PDF fits, though the smaller theory uncertainties in the beauty case provide complementary discriminating power.


Comparing the uncertainty bands for charm and beauty production, we observe that the scale uncertainty remains consistently larger than the PDF uncertainty across the entire $Q^2$ range. This dominance is also evident from the uncertainty values listed alongside the cross sections in Table~\ref{tab:heavy_quark_cross_sections}. Furthermore, the relative scale uncertainty band is noticeably narrower for beauty than for charm production. This is expected, as the larger beauty-quark mass stabilizes the cross section against variations in scale choice. It should also be kept in mind that, though the current computation is at LO, the scale uncertainty bands are expected to be reduced at NLO~\cite{Castro:2026xyr}, the current LO calculation serves our main goal of assessing the kinematic reach.

\begin{figure}[htbp!]
    \centering
    \subfloat[]{\label{fig:ep_dsig_dyc_charm}%
      \includegraphics[width=0.41\textwidth]{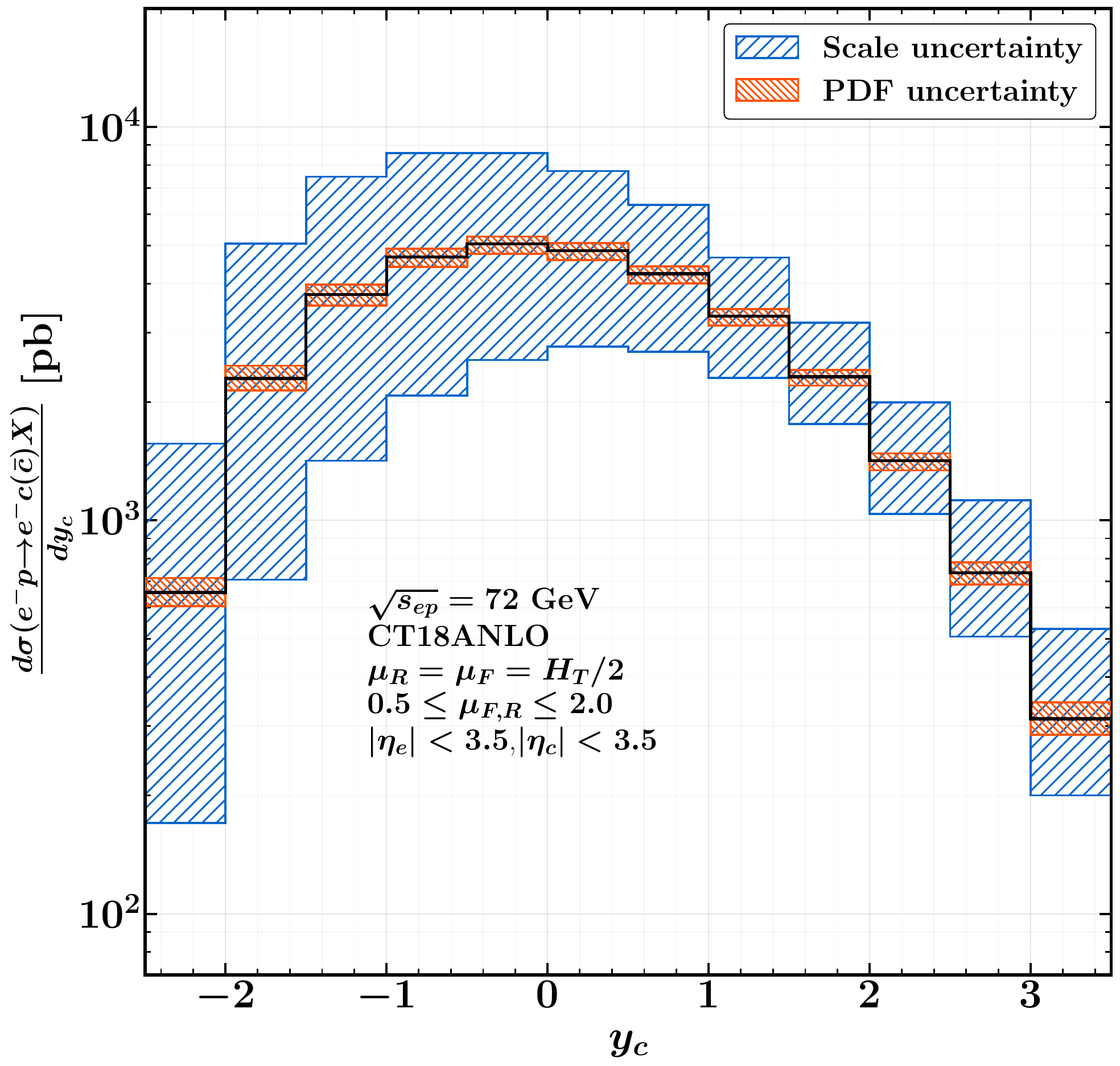}} \\
    \subfloat[]{\label{fig:ep_dsig_dyc_bottom}%
      \includegraphics[width=0.41\textwidth]{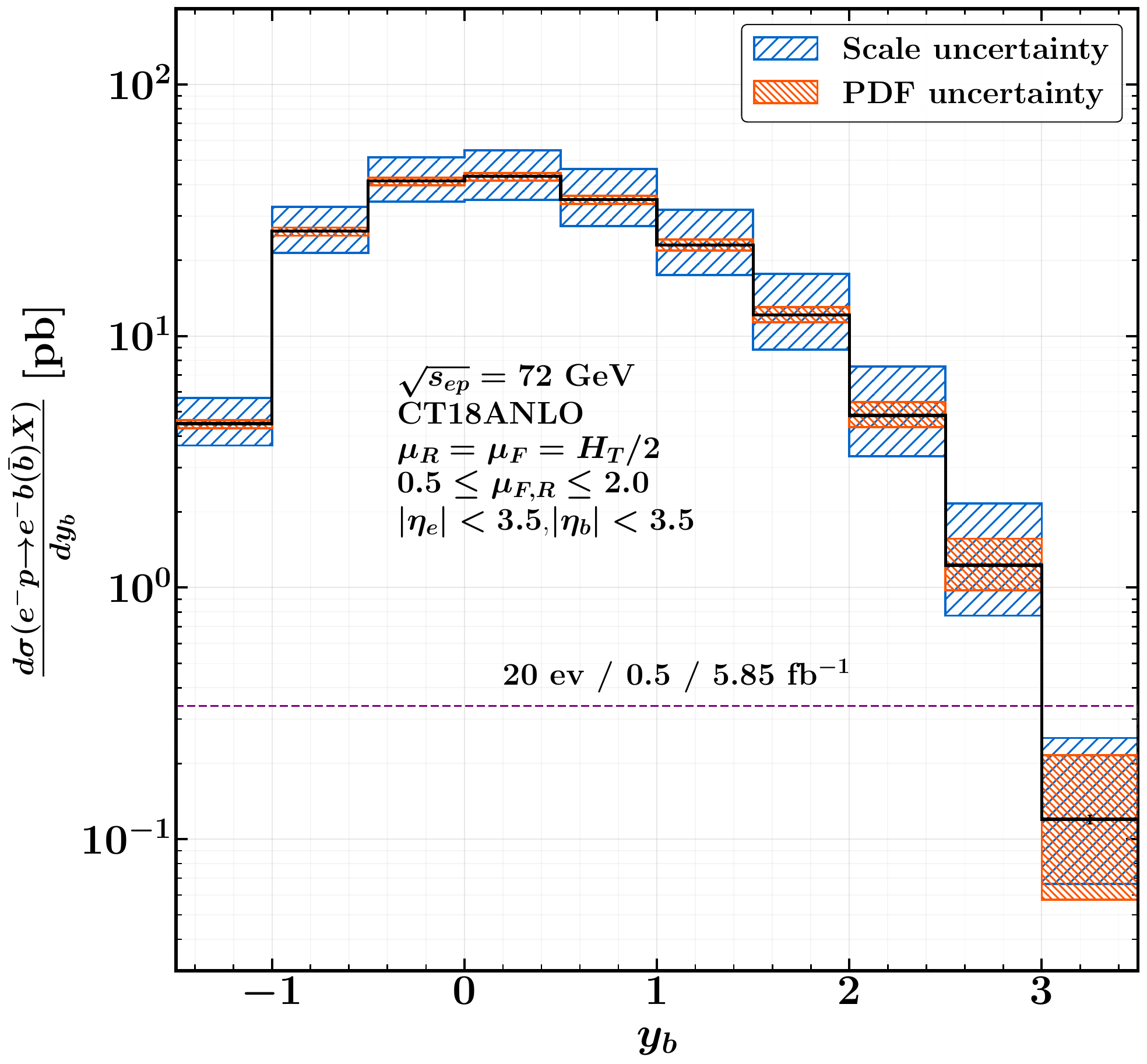}}\vspace*{-0.3cm}
    \caption{Differential cross sections in laboratory rapidity for charm $y_c$ (a) and beauty $y_b$ (b) quark production in $ep$ collisions at a c.m.~energy of $\sqrt{s_{ep}} = 72\text{ GeV}$. The hatched blue bands show the scale uncertainty, and the orange bands indicate the PDF uncertainty. The purple horizontal observability lines correspond to 20 observed events for the integrated luminosities listed in Table~\ref{tab:early-science-matrix} for each bin.}\vspace*{-0.5cm}
    \label{fig:ep_rapidity_distributions}
\end{figure}

\begin{figure}[htbp!]
\subfloat[]{\label{fig:ep_dsig_dpt_charm}%
      \includegraphics[width=0.41\textwidth]{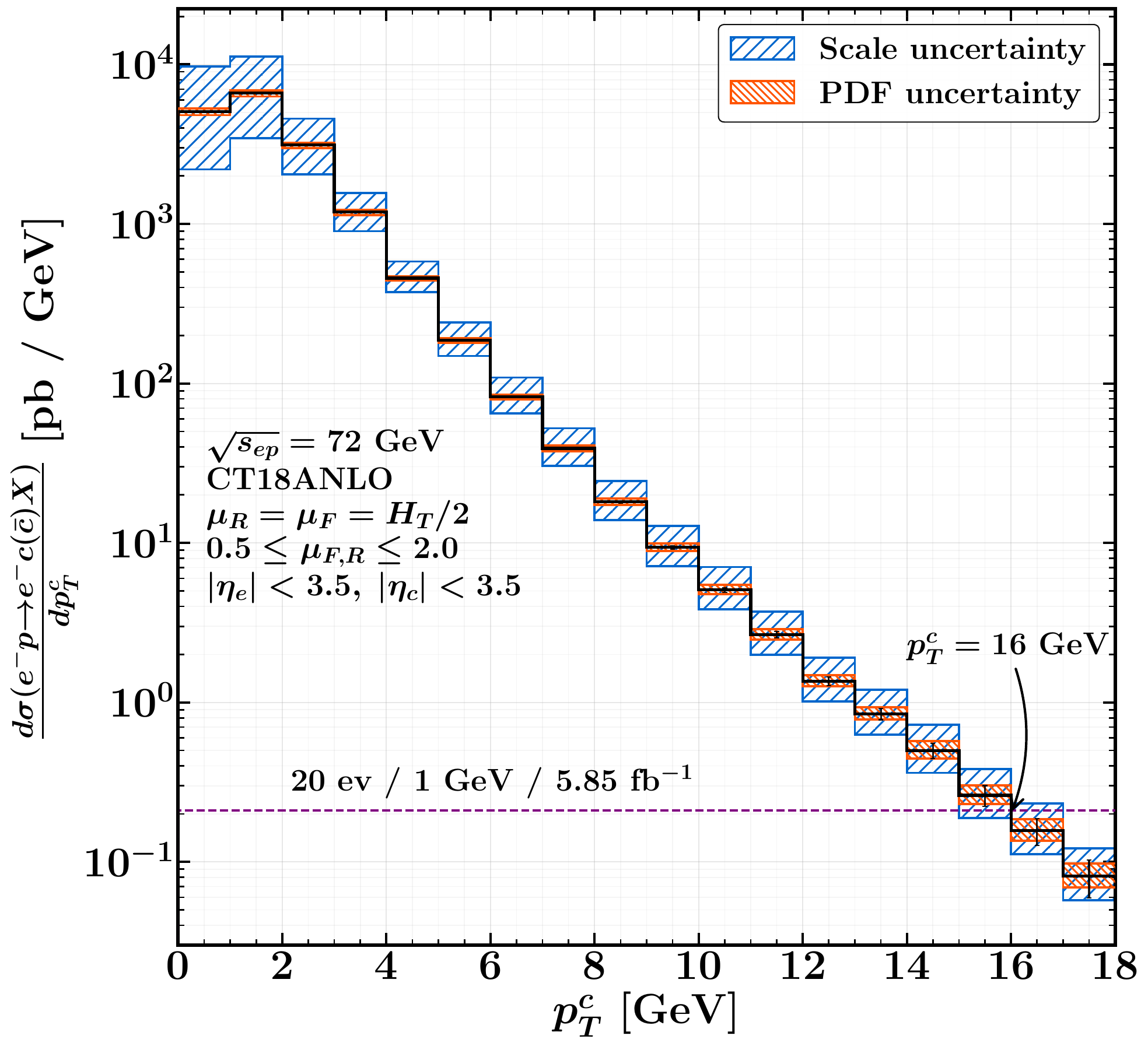}}\\
    \subfloat[]{\label{fig:ep_dsig_dpt_bottom}%
      \includegraphics[width=0.41\textwidth]{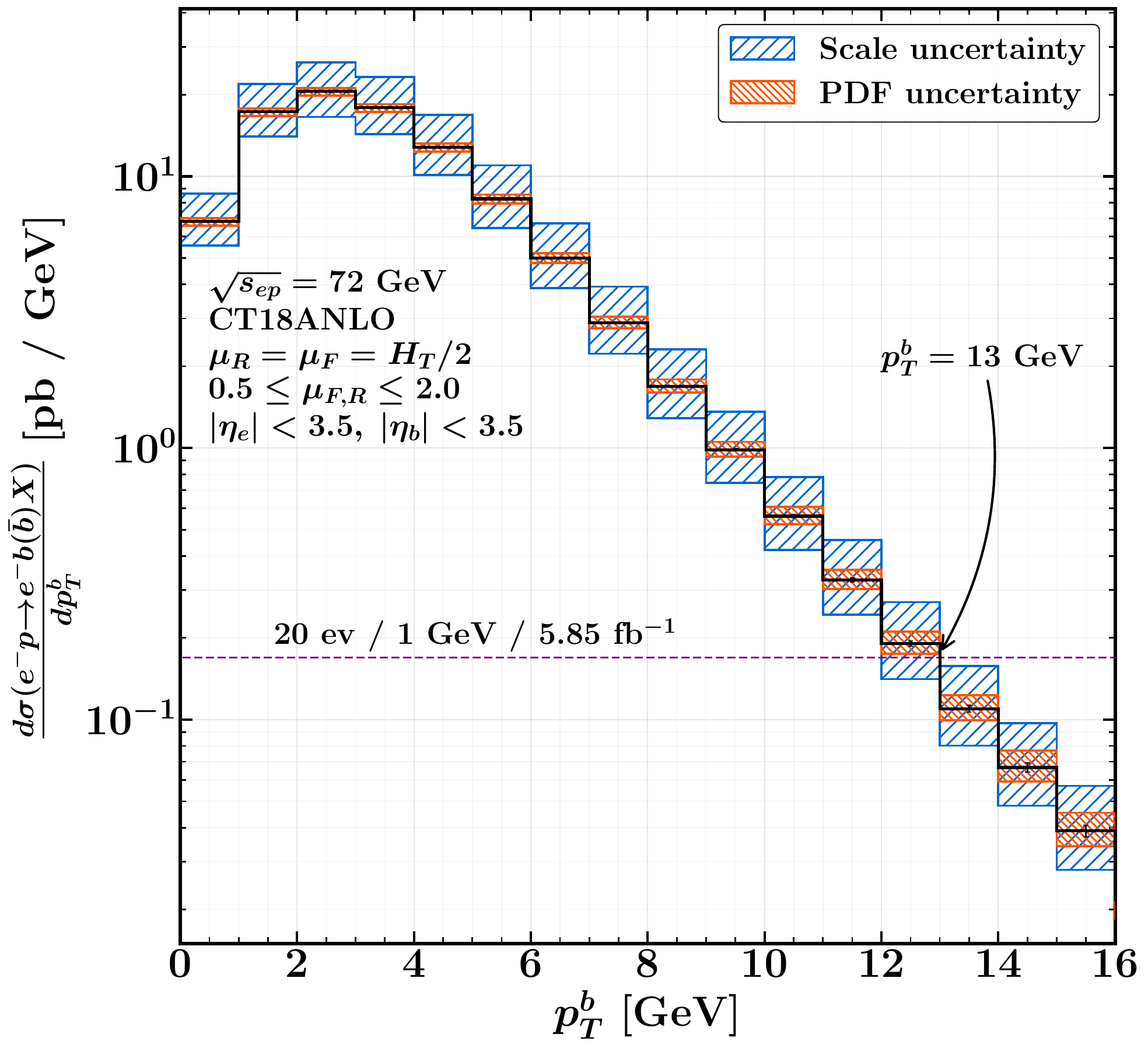}}\vspace*{-0.3cm}
    \caption{Transverse momentum cross sections $p_T^{c,b}$ for a charm (a) and beauty (b) in $ep$ collisions at $\sqrt{s_{ep}} = 72\text{ GeV}$. Hatched blue and orange bands denote scale ($0.5 \leq \mu_{F,R} \leq 2.0$) and PDF uncertainties, respectively. Purple horizontal lines indicate the 20 observed events for the luminosities listed in Table~\ref{tab:early-science-matrix}.}\vspace*{-0.5cm}
    \label{fig:ep_charm_bottom_distributions_pt}
\end{figure}

Figs.~\ref{fig:ep_dsig_dyc_charm} and \ref{fig:ep_dsig_dpt_charm} show the $y_c$\footnote{The positive rapidity is defined in the direction of the proton beam, following the same convention as in the EIC Yellow Report~\cite{AbdulKhalek:2021gbh}.} and $p_T^c$ distributions respectively of a single charm quark in $ep\to e c \bar{c} + X$ collisions. An observability line is included to indicate the kinematic reach. Since this line lies well below the populated region of the histograms, it is omitted from the $y_c$ plot. These distributions cover the rapidity range $-2.5 \leq y_c \leq 3.5$, while the $p_T^c$ spectrum extends up to $16~\text{GeV}$, corresponding to  $\langle x_{\mathrm{Bj}}\rangle \approx$ 0.30.

For beauty quarks, Figs.~\ref{fig:ep_dsig_dyc_bottom} and \ref{fig:ep_dsig_dpt_bottom} present the corresponding $y_b$ and $p_T^b$ distributions at the same centre-of-mass energy $\sqrt{s_{ep}} = 72$~GeV. The observability line shows that the full rapidity spectrum from $-1.5 \leq y_b \leq 3.0$ is accessible, while the $p_T^b$ distribution remains visible up to approximately 13 GeV, probing $\langle x_{\mathrm{Bj}}\rangle$ values as  $\approx$ 0.12. Note that the rapidity for both charm and beauty is given in the laboratory frame.

 \begin{table}[htbp!]
  \centering
  \caption{Total cross sections along with its scale and PDF uncertainty for inclusive HF electroproduction $ep \to e\,{Q}\bar{{Q}}\,X$ at $\sqrt{s_{ep}} = 72~\text{GeV}$ for $b\bar{b}$ and $c\bar{c}$ production as a function of $Q^2$. The scattered electron and heavy-quark pseudorapidities are restricted to $\vert{} \eta_{e,{Q}} \vert{} < 3.5$, consistent with the proposed ePIC detector acceptance~\cite{Li:2025lxr}.}
  \label{tab:heavy_quark_cross_sections}
  \renewcommand{\arraystretch}{1.3}
  
  \resizebox{\columnwidth}{!}{%
  \begin{tabular}{c c c c c}
    \toprule
    & \multicolumn{2}{c}{\textbf{beauty ($b\bar{b}$)}} & \multicolumn{2}{c}{\textbf{Charm ($c\bar{c}$)}} \\
    \cmidrule(lr){2-3} \cmidrule(lr){4-5}
    $Q^{2}$ [$\text{GeV}^{2}$] & $\sigma_{b\bar{b}} \,_{-\Delta_{\text{scale}}^{-}}^{+\Delta_{\text{scale}}^{+}} \,_{-\Delta_{\text{PDF}}^{-}}^{+\Delta_{\text{PDF}}^{+}}$ [pb] & $x_{{\rm Bj}}$ & $\sigma_{c\bar{c}} \,_{-\Delta_{\text{scale}}^{-}}^{+\Delta_{\text{scale}}^{+}} \,_{-\Delta_{\text{PDF}}^{-}}^{+\Delta_{\text{PDF}}^{+}}$ [pb] & $x_{{\rm Bj}}$ \\
    \midrule
    2.5--5    & $11.32 \,_{-2.35}^{+3.47} \,_{-0.40}^{+0.35}$ & $0.001 \text{ -- } 0.05$ & $2330 \,_{-920}^{+1300} \,_{-40}^{+40}$ & $0.0012 \text{ -- } 0.34$ \\
    \midrule
    5--7      & $5.18 \,_{-1.07}^{+1.59} \,_{-0.18}^{+0.16}$  & $0.002 \text{ -- } 0.07$  & $890 \,_{-320}^{+440} \,_{-13}^{+14}$    & $0.003 \text{ -- } 0.42$ \\
    \midrule
    7--12     & $7.68 \,_{-1.59}^{+2.35} \,_{-0.27}^{+0.24}$  & $0.002 \text{ -- } 0.12$   & $1090 \,_{-350}^{+470} \,_{-13}^{+16}$  & $0.003 \text{ -- } 0.55$ \\
    \midrule
    12--18    & $5.12 \,_{-1.07}^{+1.57} \,_{-0.18}^{+0.17}$  & $0.004 \text{ -- } 0.17$  & $550 \,_{-150}^{+210} \,_{-5}^{+7}$     & $0.005 \text{ -- } 0.65$ \\
    \midrule
    18--32    & $6.01 \,_{-1.26}^{+1.87} \,_{-0.22}^{+0.20}$  & $0.006 \text{ -- } 0.26$  & $470 \,_{-110}^{+150} \,_{-4}^{+4}$     & $0.009 \text{ -- } 0.77$ \\
    \midrule
    32--60    & $4.61 \,_{-0.98}^{+1.46} \,_{-0.17}^{+0.17}$ & $0.011 \text{ -- } 0.39$  & $230 \,_{-40}^{+60} \,_{-1.3}^{+1.6}$     & $0.013 \text{ -- } 0.86$ \\
    \midrule
    60--120   & $2.79 \,_{-0.61}^{+0.92} \,_{-0.11}^{+0.12}$ & $0.02 \text{ -- } 0.57$    & $80 \,_{-18}^{+20} \,_{-6}^{+8}$    & $0.024 \text{ -- } 0.92$ \\
    \midrule
    120--200$\dagger$ & $0.74 \,_{-0.32}^{+0.26} \,_{-0.04}^{+0.04}$ & $0.04 \text{ -- } 0.66$ & $20 \,_{-4}^{+6} \,_{-0.2}^{+0.4}$ & $0.036 \text{ -- } 0.95$ \\
    \midrule
    200--400 & --                                                              & --                                                           & $7 \,_{-2}^{+3} \,_{-0.1}^{+0.2}$    & $0.06 \text{ -- } 0.98$ \\
    \bottomrule
  \end{tabular}}
  \vspace{0.5em}
  \\ \raggedright \footnotesize $^{\dagger}$For beauty production, the upper range of this bin is restricted to $Q^{2} = 120\text{--}175\text{ GeV}^{2}$. 
\end{table}
 In addition to single heavy-quark electroproduction, we evaluate double-charm-pair production ($ep \to e\, c\bar{c}c\bar{c} + X$) in $ep$ collisions at $\sqrt{s_{ep}} = 72~\text{GeV}$ using the CT18ANLO PDFs. To ensure consistency, we use identical $\mu_R$ and $\mu_F$ scales alongside matching kinematic cuts on the heavy-quark pseudorapidity coverage of $|\eta_{c(\bar{c})}| < 3.5$, the scattered electron pseudorapidity of $|\eta_{e}| < 3.5$, and with an extra cut on virtuality $Q^2 > 2.5~\text{GeV}^2$. These partonic predictions are mapped to $D$ meson channels using a $c \to D^0$ fragmentation fraction.

\begin{figure}[htbp!]
    \centering
    \includegraphics[width=\columnwidth, keepaspectratio]{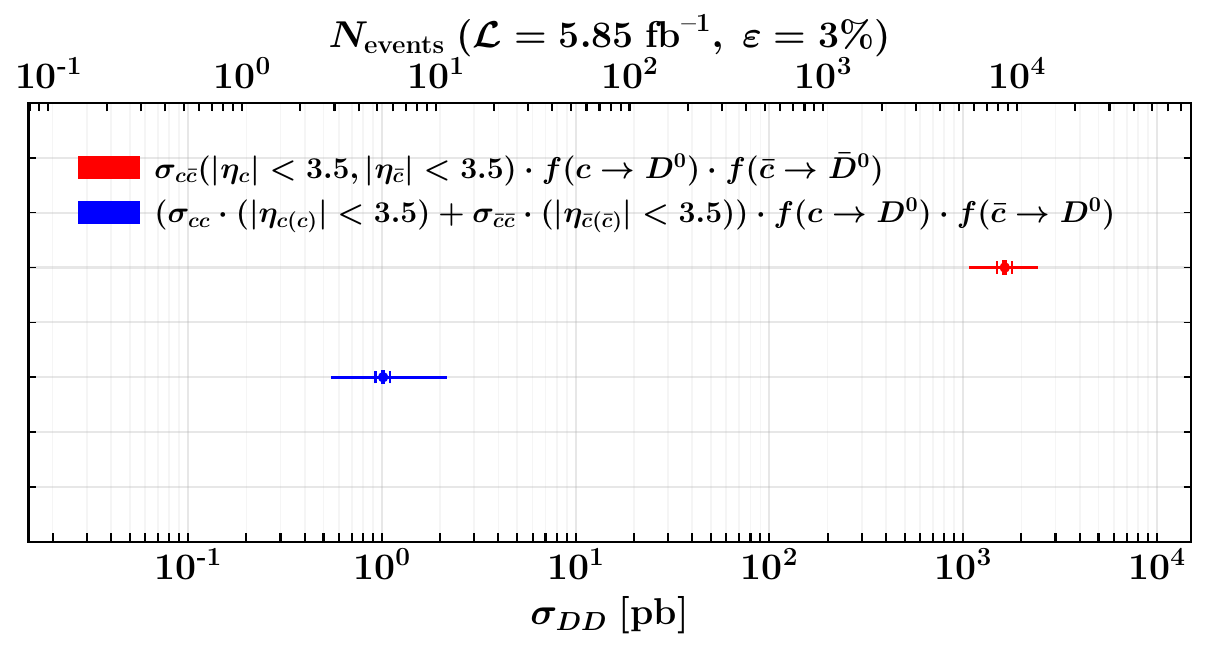}
    \caption{Predicted cross sections ($\sigma_{DD}$) and expected event yields ($N_{\text{events}}$) for opposite-sign ($D^0 \bar{D}^0$, red) and same-sign ($D^0 D^0 + \bar{D}^0 \bar{D}^0$, blue) neutral $D$ meson pair production in $ep$ collisions at $\sqrt{s} = 72\ \text{GeV}$ ($Q^2 > 2.5\ \text{GeV}^2$) with CT18ANLO PDF. Horizontal error bands for scale and PDF uncertainty are obtained from an asymmetric 9-point $\mu_R, \mu_F$ variation. The top axis maps cross sections to expected event yields assuming an integrated luminosity of $\mathcal{L} = 5.85\ \text{fb}^{-1}$ and reconstruction efficiency of $\varepsilon = 3\%$.}
    \label{fig:ep_doublecharm_total}
\end{figure}

 The opposite-sign $D^0 \bar{D}^0$ channel, driven by single $c\bar{c}$ production, yields a central cross section of $\sigma(D^0 \bar{D}^0) = 1.64_{-0.55}^{+0.76}~\text{nb}$, where the asymmetric uncertainties arise from a 9-point scale variation. In contrast, same-sign $D^0 D^0 + \bar{D}^0 \bar{D}^0$ production at LO comes from double-charm-pair production ($c\bar{c}c\bar{c}$), resulting in a significantly suppressed cross section of $1.02_{-0.46}^{+1.12}~\text{pb}$. Projecting these cross sections to expected EIC observables via $N_{\text{events}} = \sigma_{DD} \cdot \mathcal{L} \cdot \varepsilon^2$---where $\mathcal{L} = 5.85~\text{fb}^{-1}$ and $\varepsilon = 3\%$ is the $D^0$ detection efficiency, with the factor $\varepsilon^2$ accounting for both $D^0$ mesons--- we present the expected yields for each meson channel in Fig.~\ref{fig:ep_doublecharm_total}.  The opposite-sign $D^0 \bar{D}^0$ channel yields a robust signal of $N_{\text{events}} \approx 9 \times 10^3$, enabling high-precision differential studies of charm production. In contrast, the same-sign $D^0 D^0 + \bar{D}^0 \bar{D}^0$ rate is suppressed by over three orders of magnitude, producing only $N_{\text{events}} \approx 5 $ events. This indicates that observing double-charm pair production at EIC energies will be challenging and sensitive to the future detector efficiency optimizations.


\subsection{$e$A results}
We turn now to HF production in $e$A collisions at nucleon-level energies of $\sqrt{s_{eN}} = 63~\text{GeV}$ (Au, Ag, Cu) and $\sqrt{s_{eN}} = 67.8~\text{GeV}$ (Ag, Cu). These scenarios correspond to the $10\times 100~\text{GeV}$ and $10\times 115~\text{GeV}$ electron--ion beam energies scheduled for EIC operation (Table~\ref{tab:early-science-matrix}). The kinematic setup remains the same as in the $ep$ case, except for the use of the EPPS21 nuclear PDF set~\cite{Eskola:2021nhw} and the lower c.m.~energy.
Table~\ref{tab:eA_heavy_quark_cross_sections} summarises the central values together with scale and PDF uncertainties for HF production in $e$A collisions.

\begin{figure}[h!]
    \centering
    \subfloat[]{\label{fig:ep_dsig_dyc_charm_eA}
\includegraphics[width=0.41\textwidth,keepaspectratio]{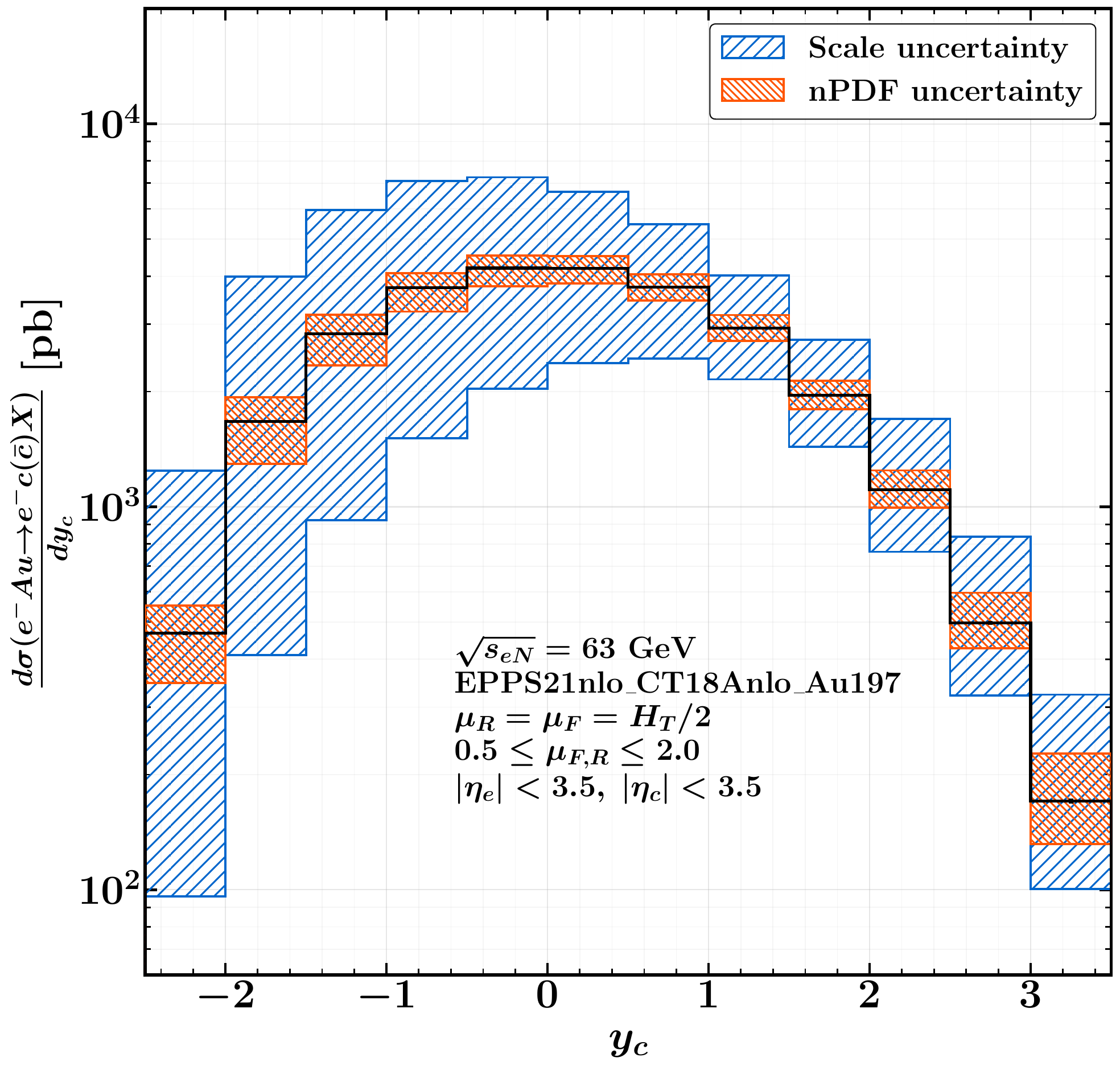}}
    \\
    \subfloat[]{\label{fig:ep_dsig_dyc_bottom_eA}
\includegraphics[width=0.41\textwidth,keepaspectratio]{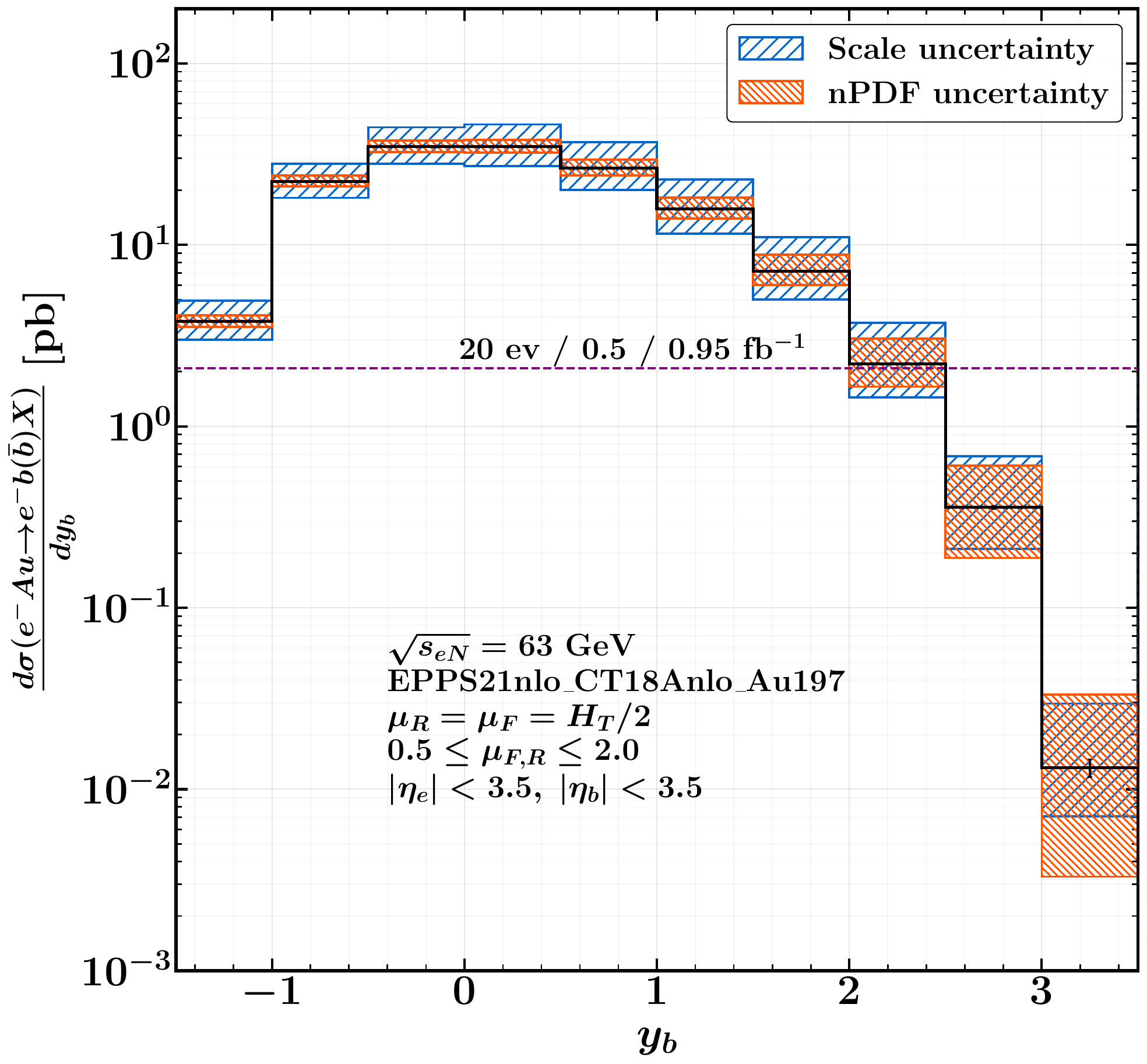}} \vspace*{-0.3cm}
\caption{Single-differential cross sections $d\sigma/dy_c$ (a) and $d\sigma/dy_b$ (b) for the inclusive production of charm  and beauty  quarks in $e$Au collisions at $\sqrt{s_{eN}} = 63\text{ GeV}$. The central renormalisation and factorisation scales are set to $\mu_R = \mu_F = H_T/2$. Kinematic acceptance cuts applied are $\vert{}\eta_e\vert{} < 3.5$ and $\vert{}\eta_{c,b}\vert{} < 3.5$.} \vspace*{-0.5cm}
    \label{fig:ep_charm_bottom_distributions_y}
\end{figure}

\begin{table}[t]
  \centering
  \caption{Total cross sections for inclusive heavy-quark electroproduction $e\text{A} \to e\,{Q}\bar{{Q}}\,X$ per nucleon in electron--nucleus collisions at the EIC for $^{197}\text{Au}$, $^{108}\text{Ag}$, and $^{64}\text{Cu}$ ions. Results are shown for centre-of-mass energies per nucleon of $\sqrt{s_{eN}} = 63~\text{GeV}$ ($10\times 100~\text{GeV}$) and $\sqrt{s_{eN}} = 67.8~\text{GeV}$ ($10\times 115~\text{GeV}$) corresponding to Year~4 operations (Table~\ref{tab:early-science-matrix}).}
  \label{tab:eA_heavy_quark_cross_sections}
  \renewcommand{\arraystretch}{1.3}
   \resizebox{\columnwidth}{!}{%
  \begin{tabular}{c c c c}
    \toprule
    Ion target & $\sqrt{s_{eN}}$ [GeV] &  $\sigma_{b\bar{b}} \,_{-\Delta_{\text{scale}}^{-}}^{+\Delta_{\text{scale}}^{+}} \,_{-\Delta_{\text{PDF}}^{-}}^{+\Delta_{\text{PDF}}^{+}}$ [pb] & $\sigma_{c\bar{c}} \,_{-\Delta_{\text{scale}}^{-}}^{+\Delta_{\text{scale}}^{+}} \,_{-\Delta_{\text{PDF}}^{-}}^{+\Delta_{\text{PDF}}^{+}}$ [pb] \\
    \midrule
    \multirow{2}{*}{\textbf{Gold ($^{197}\text{Au}$)}} & 63.0 &  73.81 $\,_{-16.6}^{+25.4} \,_{-2.8}^{+3.3}$ & 13740 $\,_{-6200}^{+9300} \,_{-800}^{+300}$ \\
    \\
    \midrule
    \multirow{2}{*}{\textbf{Silver ($^{108}\text{Ag}$)}} & 63.0 & 73.20 $\,_{-16.4}^{+25.1} \,_{-2.5}^{+3.0}$ & 13700 $\,_{-6200}^{+9300} \,_{-900}^{+400}$ \\
    & 67.8 & 86.98 $\,_{-18.8}^{+28.1} \,_{-2.9}^{+3.4}$ & 14690 $\,_{-6800}^{+10300} \,_{-900}^{+400}$ \\
    \midrule
    \multirow{2}{*}{\textbf{Copper ($^{64}\text{Cu}$)}} & 63.0 & 72.81 $\,_{-16.2}^{+24.8} \,_{-2.3}^{+2.8}$ & 13670 $\,_{-6100}^{+9300} \,_{-900}^{+400}$ \\
    & 67.8 & 86.45 $\,_{-18.6}^{+27.8} \,_{-2.7}^{+3.2}$ & 14670 $\,_{-6800}^{+10300} \,_{-900}^{+500}$ \\
    \bottomrule
  \end{tabular}
}
\end{table}
To evaluate the potential impact of the EIC on constraining nPDFs, this work focuses on observables for the gold ($^{197}\text{Au}$), copper ($^{64}\text{Cu}$) and silver ($^{108}\text{Ag}$) nuclei. Table~\ref{tab:eA_heavy_quark_cross_sections} provides the total cross sections and  uncertainties for $^{64}\text{Cu}$ and $^{108}\text{Ag}$  across two centre-of-mass energies. 
The heavy-quark rapidity distributions for Au target nuclei at $\sqrt{s_{eN}} = 63\text{ GeV}$ are presented in Figs.~\ref{fig:ep_dsig_dyc_charm_eA} and \ref{fig:ep_dsig_dyc_bottom_eA}. As indicated by the observability line, beauty production remains  accessible up to $y_b \approx 2.5$, whereas charm production is observable across the entire rapidity spectrum. Similar distributions for Ag and Cu target nuclei exhibit qualitatively identical behaviour. In this work, we examine the nuclear modification factor $R^{eA}$ 
for charm and beauty production in $e\text{A}$ collisions at $\sqrt{s_{eN}} = 63\text{ GeV}$ for all three nuclear targets: copper, silver, and gold. This ratio is evaluated differentially with respect to the heavy-quark rapidity $y_Q$ in the laboratory frame, using the CT18ANLO proton PDF set as the baseline denominator at the same collision energy.

\begin{figure}[htbp!]
    \centering
    \subfloat[]{\label{eA_dsig_dq2_charm}
    \includegraphics[width=0.41\textwidth, keepaspectratio]
    {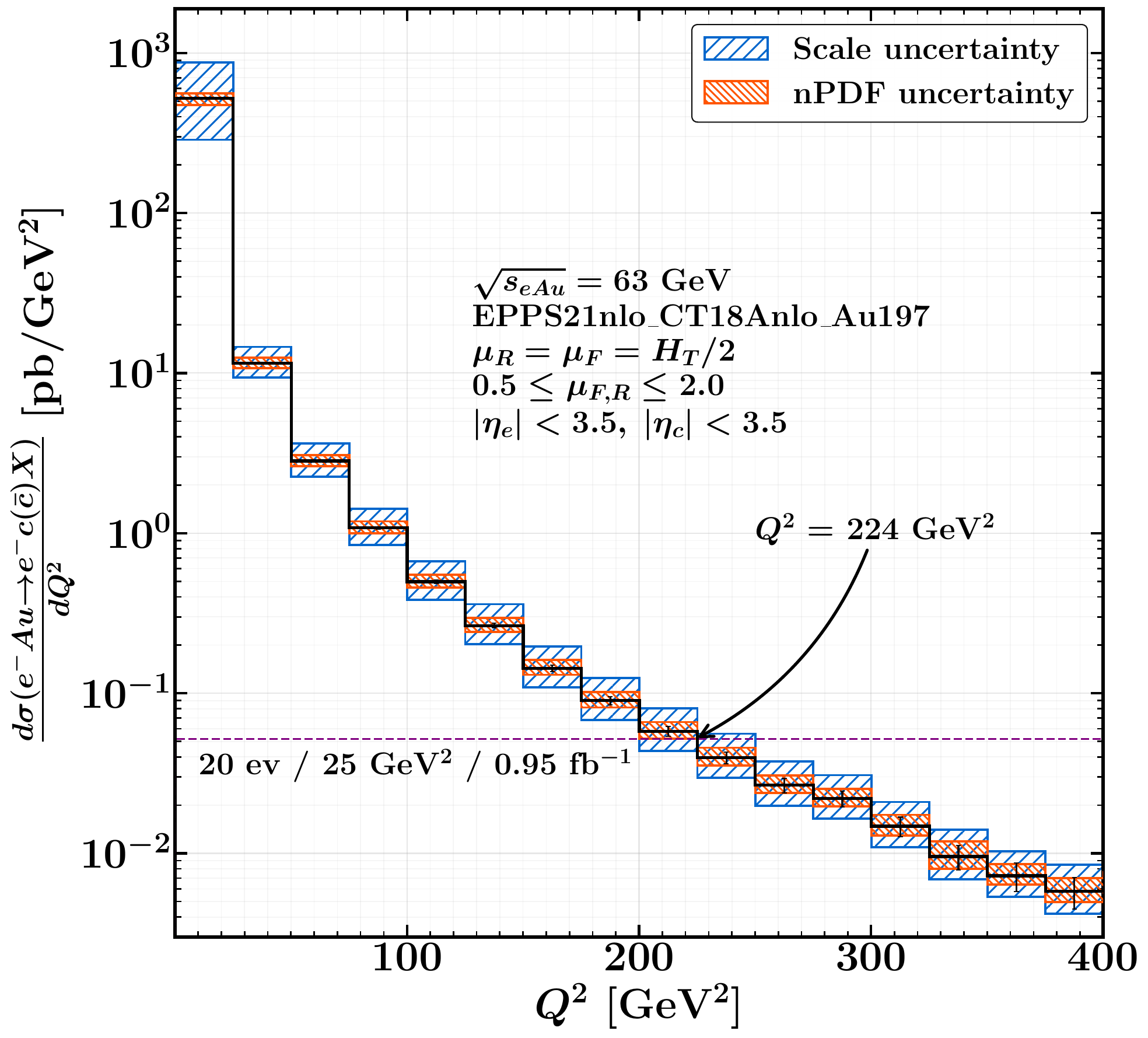}}
    \\
    \subfloat[]{\label{eA_dsig_dq2_bottom}
    \includegraphics[width=0.41\textwidth, keepaspectratio]
    {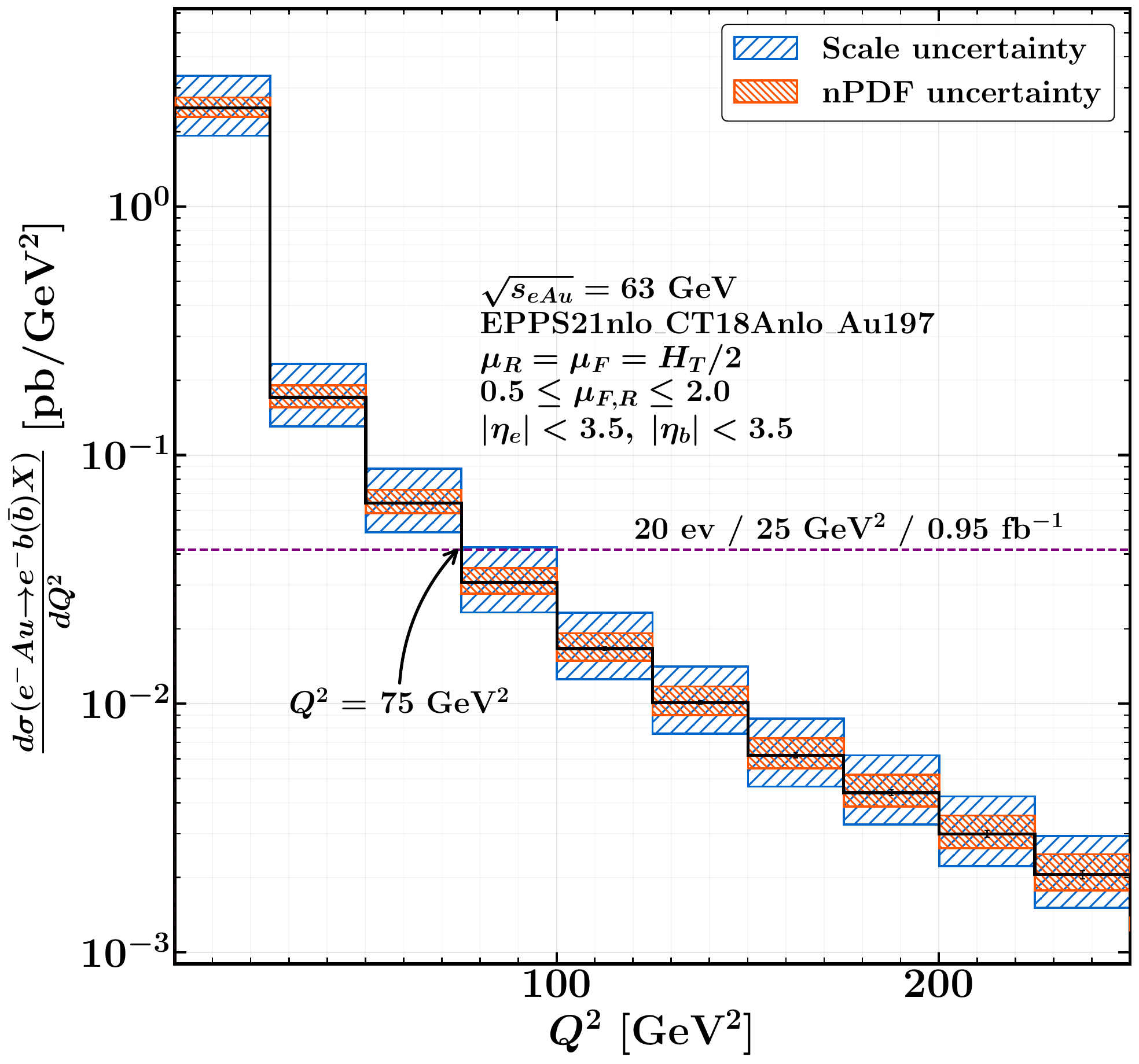}}\vspace*{-0.3cm}
    \caption{Differential cross section $\mathrm{d}\sigma/\mathrm{d}Q^2$ for $e\text{Au} \to e\, c\bar{c} + X$ (a)
    and $e\text{Au} \to e\, b\bar{b} + X$ (b)
    at $\sqrt{s_{eN}} = 63$~GeV. The blue hatched bands represent the scale uncertainty. The purple horizontal histogram indicates the observability threshold corresponding to the projected Year~4 eAu luminosity of $0.95~\mathrm{fb}^{-1}/\mathrm{nucleon}$~\cite{Arleo:2026tpb}}. \vspace*{-0.5cm}
    \label{fig:eA_dsig_dq2}
\end{figure}

The differential cross sections $\mathrm{d}\sigma/\mathrm{d}Q^2$ distributions for charm and beauty production in $e$Au are shown in Figs.~\ref{eA_dsig_dq2_charm} and~\ref{eA_dsig_dq2_bottom} respectively. Charm production remains visible up to $Q^2 \approx 224~\text{GeV}^2$ (Fig.~\ref{eA_dsig_dq2_charm})  probing up to $\langle x_{\rm{Bj}}\rangle \approx 0.66$, whereas beauty is visible only up to $Q^2 \approx 75~\text{GeV}^2$ (Fig.~\ref{eA_dsig_dq2_bottom}), reaching $\langle x_{\rm{Bj}}\rangle \approx$ 0.25. This significantly broader $Q^2$ reach for charm reflects the stronger mass suppression experienced by the heavier beauty quark at $\sqrt{s_{eN}} = 63~\text{GeV}$.

\begin{figure}[htbp!]
    \centering
    \subfloat[]{\label{eAu_dsig_dx_charm}
        \centering
        \includegraphics[width=0.35\textwidth]{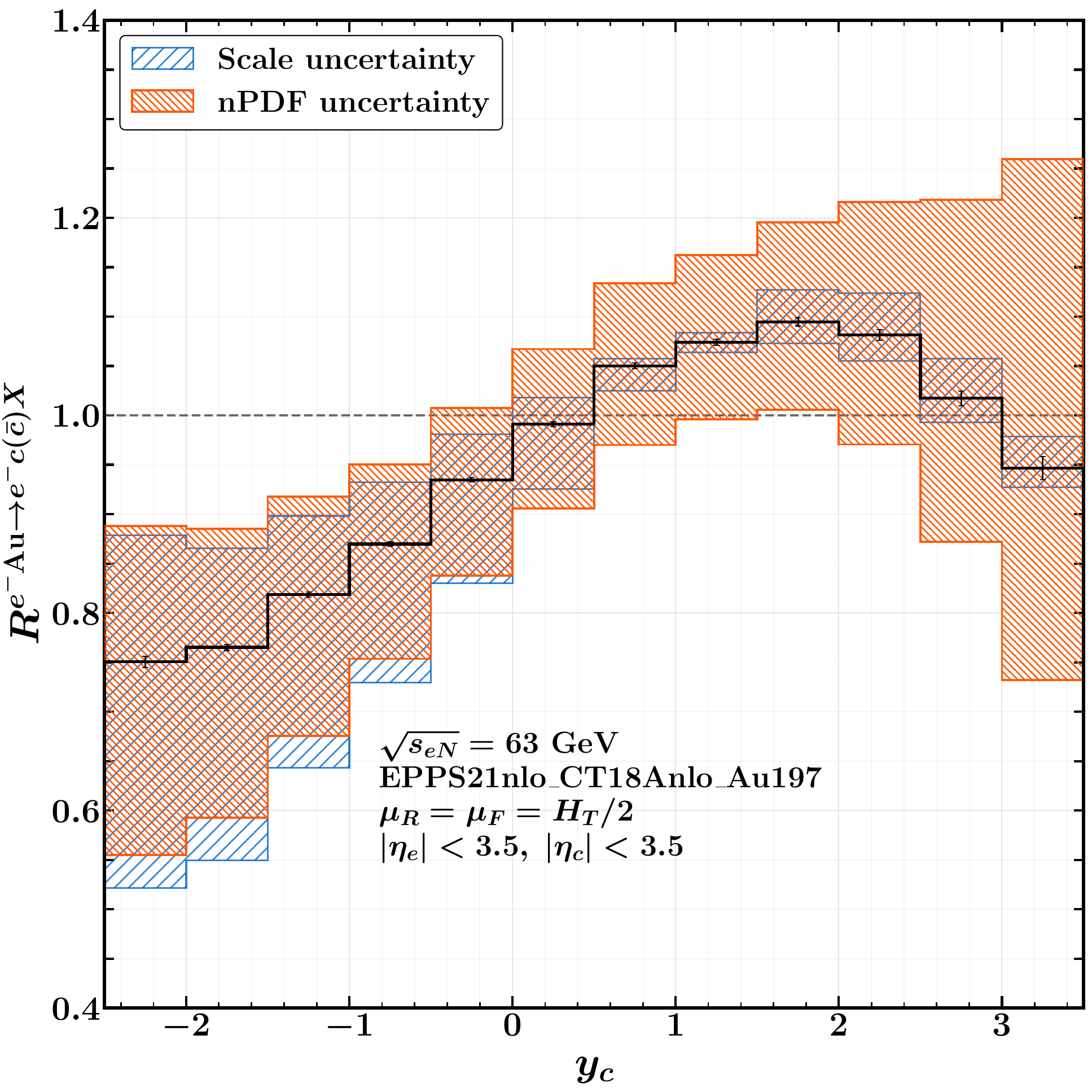}} \vspace*{-0.3cm}
    \\
    \subfloat[]{\label{eAg_dsig_dx_charm}
        \centering
        \includegraphics[width=0.35\textwidth]{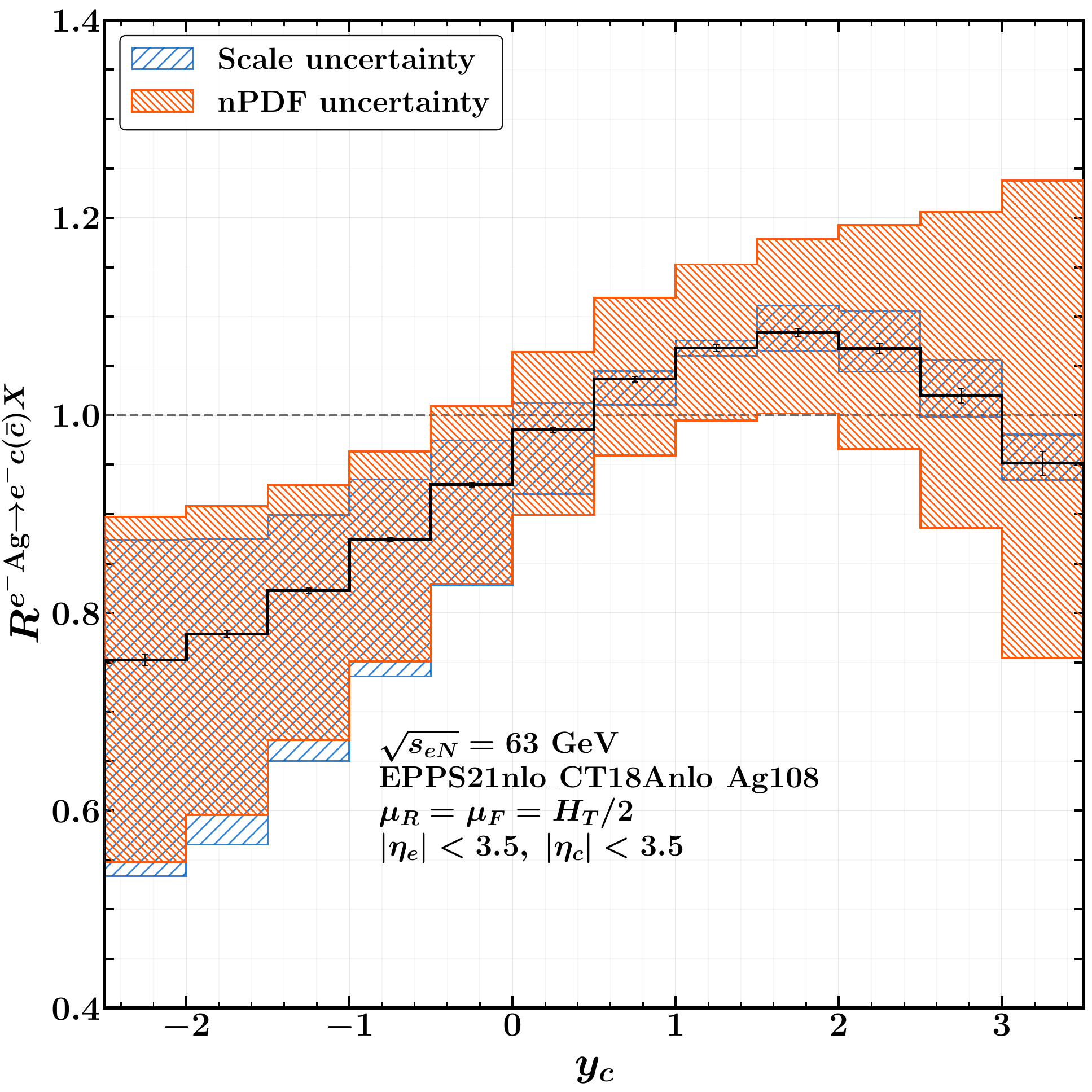}}\vspace*{-0.3cm}
    \\
   \subfloat[]{\label{eCu_dsig_dx_charm}
        \centering
        \includegraphics[width=0.35\textwidth]{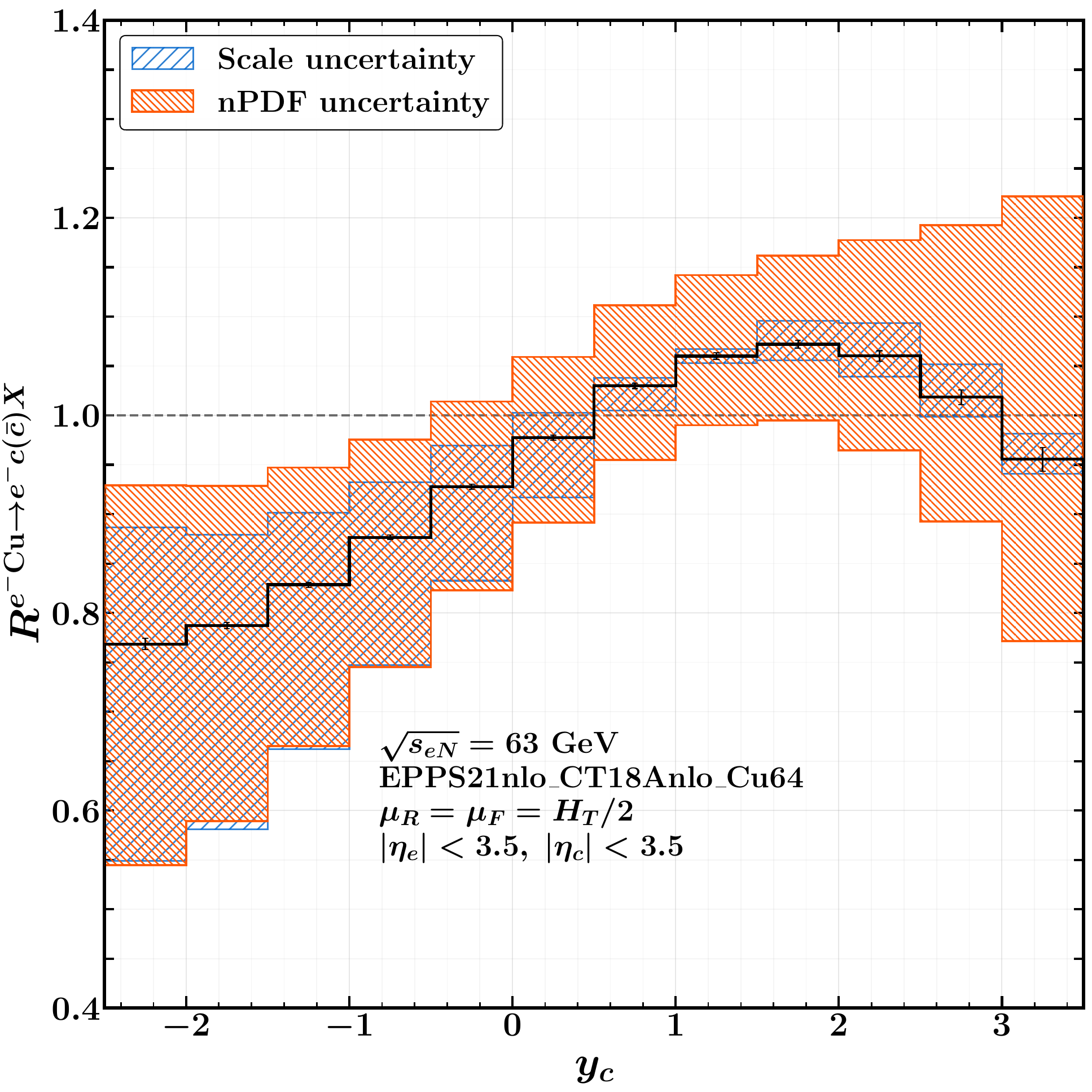}}
    \caption{Nuclear modification ratio $R^{e\text{A}}$ for charm production across three nuclear targets (Au (a), Ag (b) and Cu (c)) as a function of rapidity at $\sqrt{s_{eN}} = 63$~GeV. The ratios are computed using EPPS21nlo nPDFs~\cite{Eskola:2021nhw} normalized to the CT18ANLO proton baseline. The shaded bands represent the corresponding uncertainty envelopes for each nucleus.}
    \label{eA_dsig_dx_charm}
\end{figure}

\begin{figure}[htbp!]
    \centering
    \subfloat[]{\label{eAu_dsig_dx_bottom}
        \centering
        \includegraphics[width=0.35\textwidth]{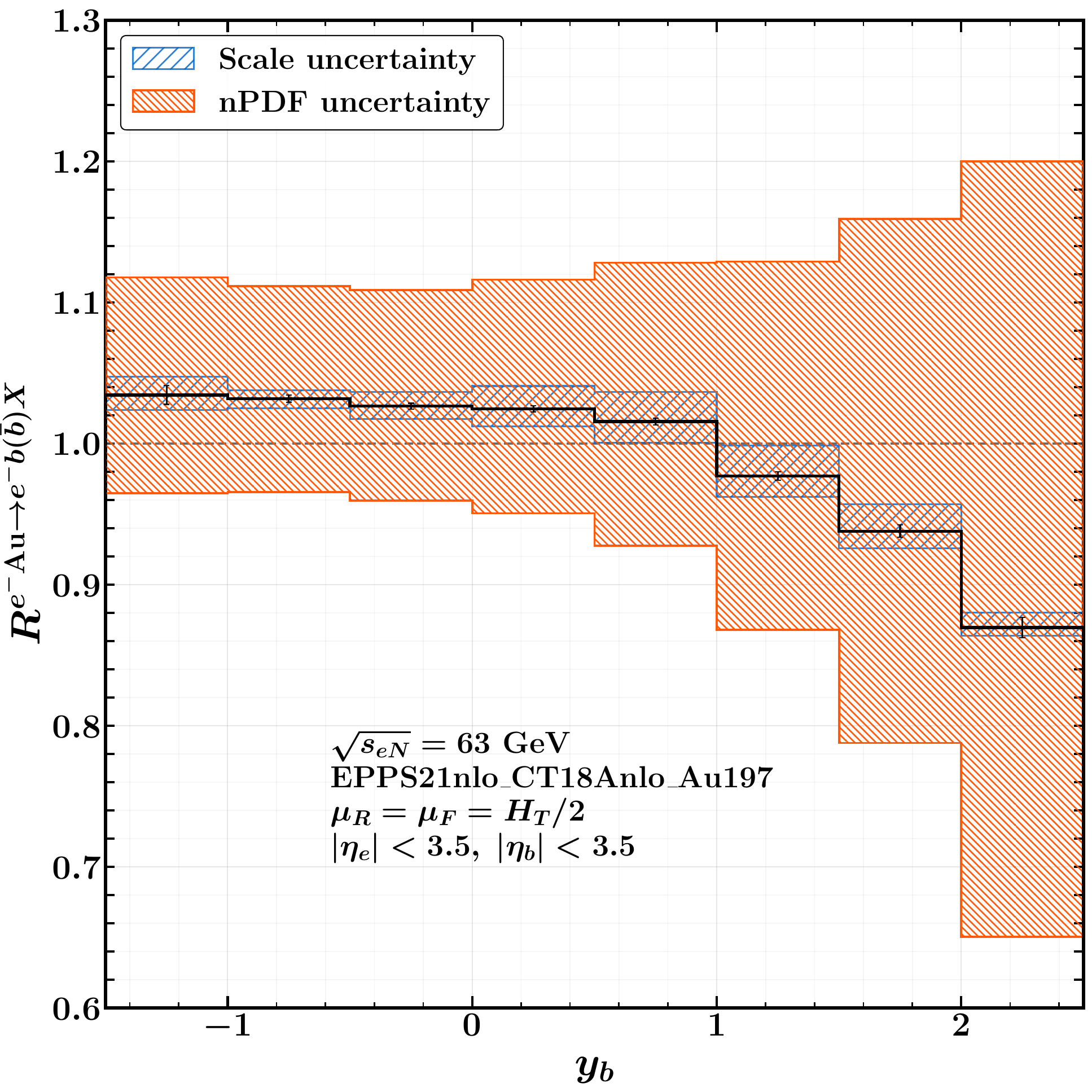}} \vspace*{-0.3cm}
    \\
    \subfloat[]{\label{eAg_dsig_dx_bottom}
        \centering
        \includegraphics[width=0.35\textwidth]{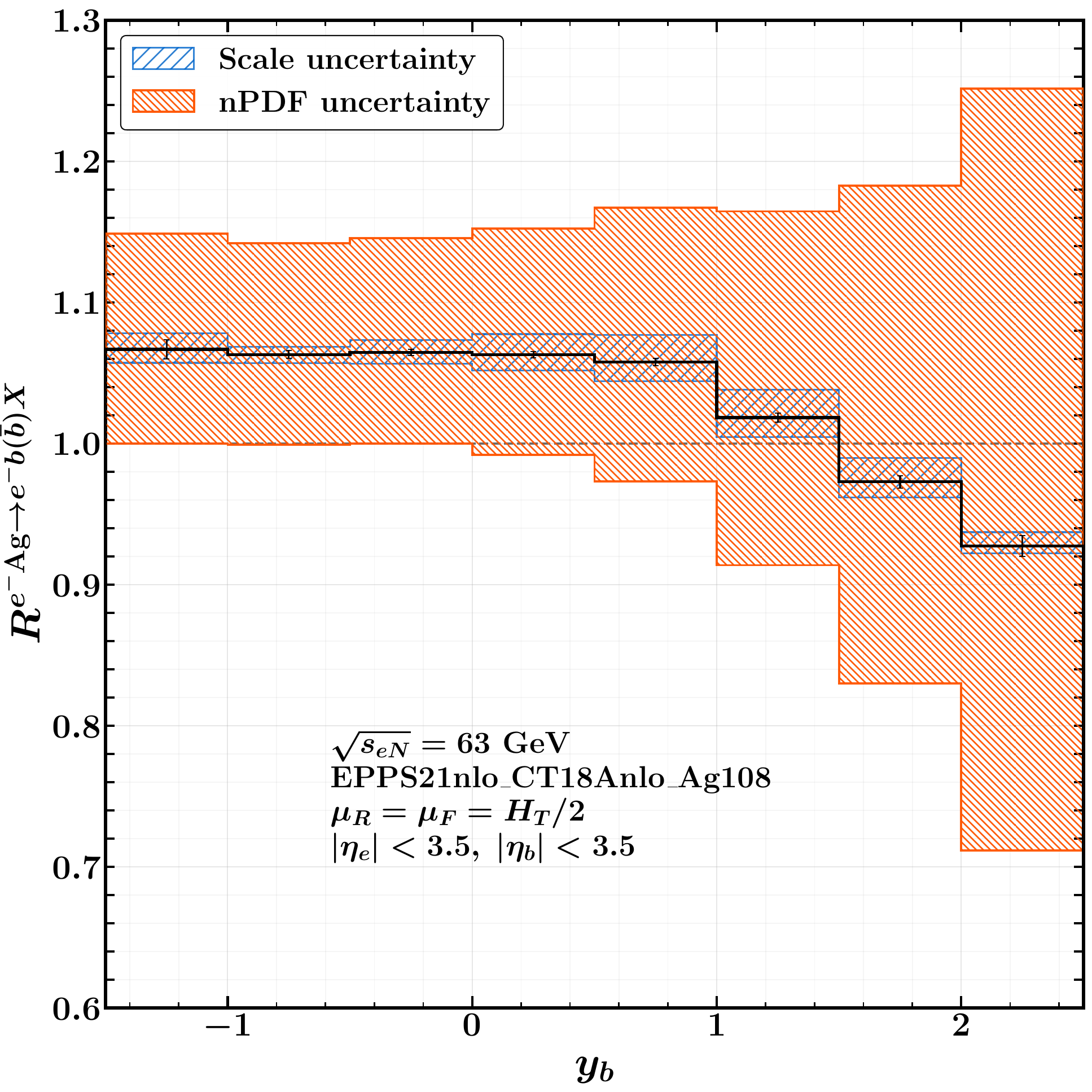}}\vspace*{-0.3cm}
    \\
   \subfloat[]{\label{eCu_dsig_dx_bottom}
        \centering
        \includegraphics[width=0.35\textwidth]{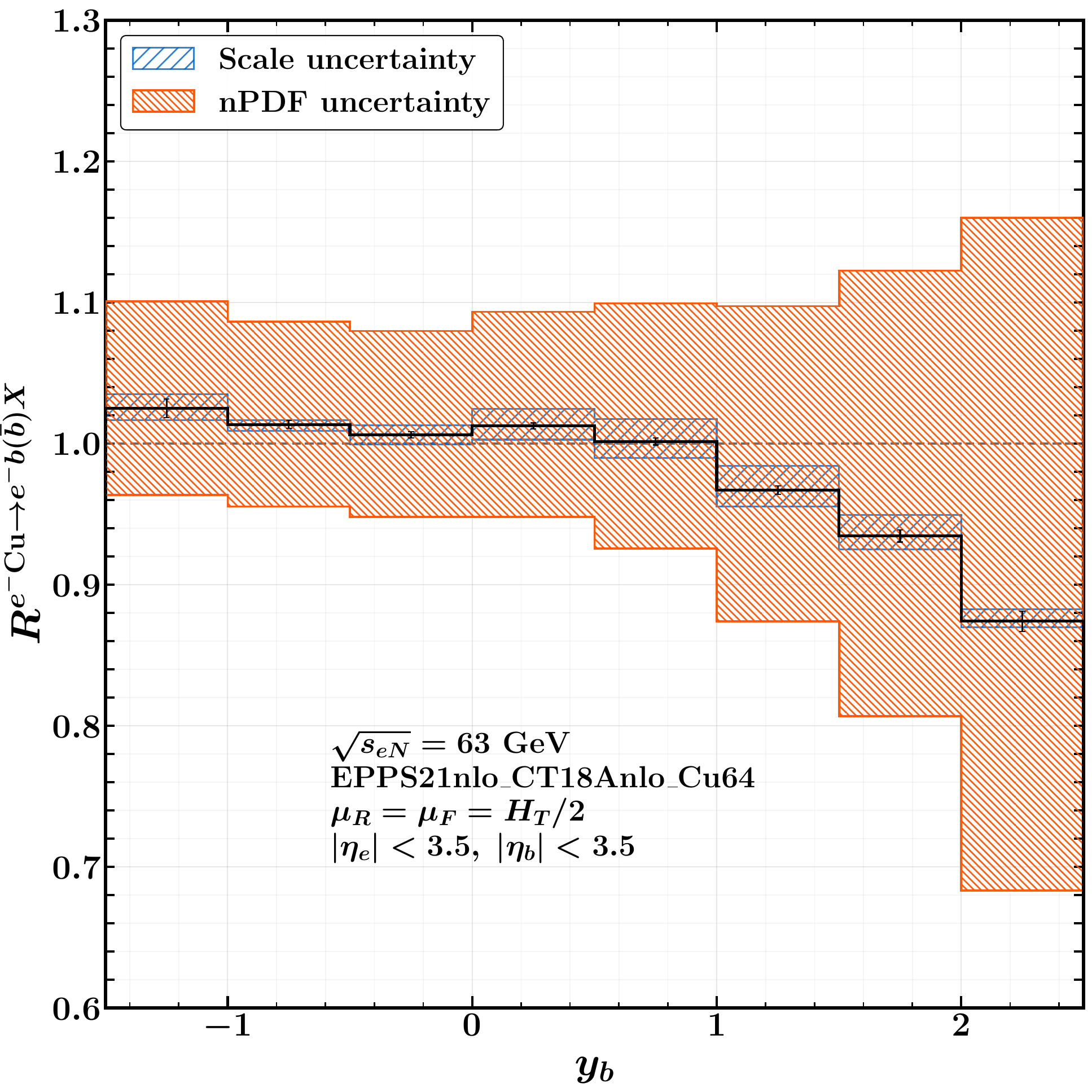}}\vspace*{-0.3cm}
    \caption{Nuclear modification ratio $R^{eA}$ for beauty production across three nuclear targets ($\mathrm{Au}$ (a), $\mathrm{Ag}$ (b), and $\mathrm{Cu}$ (c)) as a function of rapidity $y_b$ at $\sqrt{s_{eN}} = 63$~GeV. The ratios are computed using EPPS21nlo nPDFs~\cite{Eskola:2021nhw} normalized to the CT18ANLO proton baseline. 
    }
    \label{eA_dsig_dx_bottom}
\end{figure}

The nuclear modification factors are shown as functions of rapidity for charm and beauty production in Fig.~\ref{eA_dsig_dx_charm} and  Fig.~\ref{eA_dsig_dx_bottom} respectively, computed using EPPS21nlo PDFs~\cite{Eskola:2021nhw,Eskola:2009uj}. Following the EIC convention, where the nucleus travels in the $+z$ direction, the gluon momentum fraction is related to the heavy-quark rapidity such that forward rapidity corresponds to larger $x_g$\footnote{Momentum fraction carried by the gluon.} and backward rapidity to smaller $x_g$.

For charm production (Fig.~\ref{eA_dsig_dx_charm}), the nuclear modification ratio $R^{eA}$ across all three ions exhibits a pronounced rapidity dependence alongside a clear atomic mass number ($A$) ordering. At the most backward rapidities ($y_c \lesssim -1$), where small $x_g$ is probed, the ratio is strongly suppressed below unity, reaching $R^{eCu}\approx0.78$ for copper and $R^{eAu}\approx0.75$ for gold. This is consistent with gluon shadowing at small $x_g$. Moving to central and forward rapidities ($y_c \approx 0.5-2.0$), the ratios rise to anti-shadowing maxima ranging from 1.07 for copper to 1.10 for gold, before turning downward at the most forward rapidities ($y_c > 2.5$), where large $x_g$ is probed. Here, EMC-like suppression seems to set in, though a more definitive conclusion would require higher-order calculations and experimental input since the nPDF uncertainties here are large.

For beauty quark production (Fig.~\ref{eA_dsig_dx_bottom}), the significantly larger quark mass ($m_b \approx 4.75\text{ GeV}$) shifts the accessible momentum fraction $x_g$ toward larger values at equivalent rapidities 
compared to charm. Across backward and central rapidities ($y_b \lesssim 1.0$), corresponding to intermediate-to-large $x_g$, the nuclear modification ratios $R^{eA}$ for all three targets ($\mathrm{Au}$, $\mathrm{Ag}$, and $\mathrm{Cu}$) remain close to unity ($R^{eA} \approx 1.00\text{--}1.06$), reflecting an anti-shadowing region. 
Moving toward forward rapidities ($y_b \gtrsim 2.0$), where $x_g$ is larger, $R^{eA}$ exhibits a mild decrease down to $R^{eA} \approx 0.87\text{--}0.93$, consistent with EMC-like suppression at large $x_g$. In this far-forward region, the nPDF uncertainty bands (orange) expand dramatically for all ions, reflecting the large theoretical uncertainties in nuclear gluon distributions at large $x_g$ and thus the importance of such EIC measurements.

\section{Conclusion and Outlook}\label{sec:outlook}

We have presented a study of charm and beauty quark electroproduction in electron-proton ($ep$) and electron-nucleus ($e$A) collisions at centre-of-mass energies of $\sqrt{s_{ep}} = 72~\mathrm{GeV}$ and $\sqrt{s_{eN}} = 63\ \& \ 67.8~\mathrm{GeV}$ respectively, corresponding to the beam configurations planned for the early years of EIC operation. Using the CT18ANLO PDF set at leading order, we computed differential cross sections as functions of $Q^2$, heavy-quark rapidity $y_{c/b}$, and transverse momentum $p_T^{c/b}$, and estimated the kinematic regions accessible with the projected integrated luminosities via observability lines.

In $ep$ collisions, charm production yielded cross sections of several hundreds of pb, remaining observable over a broad range of $Q^2$, rapidity, and $x_{\rm{Bj}}$, while beauty production is suppressed relative to charm by the larger quark mass, but remains accessible, particularly in the central and forward rapidity regions. Taken together, these results establish a parton-level baseline for early EIC HF measurements and demonstrate the substantial improvement in statistical reach relative to the HERA programme. Moreover, we showed that unlike-sign charm pairs will be readily observable with $\mathcal{O}(10^3)$ events, enabling differential studies of charm fragmentation and photon-gluon fusion dynamics, 
while like-sign pairs -- though a clean probe of double-charm-pair production and a valuable constraint on double-parton-scattering -- remain statistically challenging, but potentially measurable as early as the first year of operation —- a capability that was beyond the reach of HERA.

Turning to $e$A collisions, we employed the CT18ANLO proton PDF 
together with the EPPS21 nuclear PDF set to study nuclear 
modifications of the HF yield. The nuclear modification 
factor $R^{e\text{A}}$ exhibits a rich structure across the accessible 
rapidity range: gluon shadowing at low $x_g$, anti-shadowing at 
intermediate $x_g$, and EMC-like suppression at higher $x_g$. 
Due to the mass of the beauty being heavier than charm, beauty production probes larger values of $x_g$ at equivalent rapidities, and therefore does not access the small-$x_g$ shadowing regime as deeply as charm, remaining instead in the anti-shadowing and EMC regions across the accessible rapidity range.  
These mass-dependent differences highlight the 
complementary sensitivity of the two heavy flavours to distinct 
nuclear regimes and underscore the value of measuring both. While we have evaluated $R^{e\text{A}}$ for gold, silver, and copper ions, this analysis can be easily extended to other nuclei thanks to \texttt{MadGraph5\_aMC@NLO}. Such computations can also be reproduced online thanks to the Virtual Access NLOAccess~\cite{Flore:2023dps} at \url{https://nloaccess.in2p3.fr/} and could be extended to quarkonium studies thanks to the new extension of \texttt{MadGraph5\_aMC@NLO}~\cite{ColpaniSerri:2025vdz,Maxia:2026ved}
as it was already done in~\cite{Boer:2024ylx} with \texttt{HELAC-Onia}~\cite{Shao:2012iz,Shao:2015vga}. Extensions to photoproduction (up to NLO) will soon be available including resolved-photon contributions~\cite{Manna:2024ltm} thanks to the asymmetric version of \texttt{MadGraph5\_aMC@NLO}~\cite{Flore:2025ync}.

These results demonstrate that heavy-quark-pair electroproduction at the EIC will provide a powerful and theoretically clean probe of nuclear parton distribution functions, with particular sensitivity to the gluon density at small $x_g$. The simultaneous availability of $ep$ and $e$A data, combined with the EIC's high luminosity and broad kinematic coverage, offers a uniquely controlled environment for mapping nuclear modifications with implications for the interpretation of heavy-ion data at RHIC and the LHC. Furthermore, precise heavy-flavour baselines in unpolarised DIS will serve as an indispensable prerequisite for disentangling nonperturbative QCD effects—most notably intrinsic charm in the nucleon~\cite{Brodsky:1980pb,Blumlein:2015qcn,Hou:2017khm,Ball:2022intrinsic}—and for guiding dedicated experimental searches at the EIC~\cite{Courtoy:2023intrinsic}.    This physics case is further strengthened by associated $J/\psi + c$ production at the EIC, which provides a complementary handle on intrinsic charm~\cite{Flore:2020jau} and underscores the need for the precise open heavy-flavour baselines established here.

\section*{Acknowledgments}
We thank C. Flore, D. Kiko\l a, K. Lynch, and C. Van Hulse
for useful discussions and inputs.

\noindent
This publication has emanated from research
conducted with the financial support of Taighde
\'Eireann -- Research Ireland under Grant number GOIPG/2023/5087.

This work has also been supported in part by the Excellence Initiative: Research University at Warsaw
University of Technology and the European Union’s Horizon 2020 research Grant Agreements no. 722104 as a part of the Marie Skłodowska-Curie Innovative Training Network
MCnetITN3.

\bibliographystyle{utphys}

\bibliography{bib}

@article{Manna:2024ltm,
    author = "Manna, Laboni and Safronov, Anton and Flore, Carlo and Kikola, Daniel and Lansberg, Jean-Philippe and Mattelaer, Olivier",
    title = "{Resolved photoproduction in MadGraph5{\_}aMC@NLO}",
    eprint = "2410.17061",
    archivePrefix = "arXiv",
    primaryClass = "hep-ph",
    doi = "10.22323/1.469.0185",
    journal = "PoS",
    volume = "DIS2024",
    pages = "185",
    year = "2025"
}

@article{Flore:2025ync,
    author = "Flore, Carlo and Kiko{\l}a, Daniel and Kusina, Aleksander and Lansberg, Jean-Philippe and Mattelaer, Olivier and Safronov, Anton",
    title = "{Automated NLO calculations for asymmetric hadron-hadron collisions in MadGraph5{\_}aMC@NLO}",
    eprint = "2501.14487",
    archivePrefix = "arXiv",
    primaryClass = "hep-ph",
    doi = "10.1140/epja/s10050-025-01707-1",
    journal = "Eur. Phys. J. A",
    volume = "61",
    number = "10",
    pages = "239",
    year = "2025"
}

@article{Shao:2015vga,
    author = "Shao, Hua-Sheng",
    title = "{HELAC-Onia 2.0: an upgraded matrix-element and event generator for heavy quarkonium physics}",
    eprint = "1507.03435",
    archivePrefix = "arXiv",
    primaryClass = "hep-ph",
    reportNumber = "CERN-PH-TH-2015-155",
    doi = "10.1016/j.cpc.2015.09.011",
    journal = "Comput. Phys. Commun.",
    volume = "198",
    pages = "238--259",
    year = "2016"
}

@article{Shao:2012iz,
    author = "Shao, Hua-Sheng",
    title = "{HELAC-Onia: An automatic matrix element generator for heavy quarkonium physics}",
    eprint = "1212.5293",
    archivePrefix = "arXiv",
    primaryClass = "hep-ph",
    doi = "10.1016/j.cpc.2013.05.023",
    journal = "Comput. Phys. Commun.",
    volume = "184",
    pages = "2562--2570",
    year = "2013"
}

@article{Boer:2024ylx,
    author = {Boer, Dani{\"e}l and others},
    title = "{Physics case for quarkonium studies at the Electron Ion Collider}",
    eprint = "2409.03691",
    archivePrefix = "arXiv",
    primaryClass = "hep-ph",
    doi = "10.1016/j.ppnp.2025.104162",
    journal = "Prog. Part. Nucl. Phys.",
    volume = "142",
    pages = "104162",
    year = "2025"
}

@article{Maxia:2026ved,
    author = "Maxia, Luca and Shao, Hua-Sheng and Simon, Lukas",
    title = "{Automated NRQCD and NRQED simulations of quarkonium and leptonium production with P-wave states and physical-mass effects}",
    eprint = "2607.26739",
    archivePrefix = "arXiv",
    primaryClass = "hep-ph",
    month = "7",
    year = "2026"
}

@article{ColpaniSerri:2025vdz,
    author = "Colpani Serri, Alice and Flett, Chris A. and Lansberg, Jean-Philippe and Mattelaer, Olivier and Shao, Hua-Sheng and Simon, Lukas",
    title = "{Automated event generation for S-wave quarkonium and leptonium production in NRQCD and NRQED}",
    eprint = "2510.26773",
    archivePrefix = "arXiv",
    primaryClass = "hep-ph",
    doi = "10.1007/JHEP02(2026)159",
    journal = "JHEP",
    volume = "02",
    pages = "159",
    year = "2026"
}

@article{Flore:2023dps,
    author = "Flore, Carlo",
    title = "{NLOAccess: automated online computations for collider physics}",
    eprint = "2301.09167",
    archivePrefix = "arXiv",
    primaryClass = "hep-ph",
    doi = "10.1140/epja/s10050-023-00972-2",
    journal = "Eur. Phys. J. A",
    volume = "59",
    number = "3",
    pages = "46",
    year = "2023"
}

@article{Anderle:2021eicc,
    author        = "Anderle, D. P. and others",
    title         = "{Electron-ion collider in China}",
    eprint        = "2103.01865",
    archivePrefix = "arXiv",
    primaryClass  = "hep-ph",
    journal       = "Front. Phys. (Beijing)",
    volume        = "16",
    number        = "6",
    pages         = "64701",
    year          = "2021"
}

@article{Ball:2024vfns,
    author        = "Ball, Richard D. and others",
    title         = "{Heavy quark mass effects in DIS structure functions}",
    eprint        = "2408.07383",
    archivePrefix = "arXiv",
    primaryClass  = "hep-ph",
    year          = "2024"
}

@article{Ball:2022intrinsic,
    author        = "Ball, Richard D. and others",
    collaboration = "NNPDF",
    title         = "{Evidence for intrinsic charm in the proton}",
    eprint        = "2208.08372",
    archivePrefix = "arXiv",
    primaryClass  = "hep-ph",
    journal       = "Nature",
    volume        = "608",
    number        = "7924",
    pages         = "483--487",
    year          = "2022"
}

@article{Courtoy:2023intrinsic,
    author        = "Courtoy, A. and Garzelli, M. V. and Olness, F. I. and Kusina, A. and Schienbein, I. and Yu, J. Y.",
    title         = "{Probing intrinsic charm at the EIC}",
    eprint        = "2311.00743",
    archivePrefix = "arXiv",
    primaryClass  = "hep-ph",
    year          = "2023"
}

@article{Andronic:2015wma,
    author = "Andronic, A. and others",
    title = "{Heavy-flavour and quarkonium production in the LHC era: from proton{\textendash}proton to heavy-ion collisions}",
    eprint = "1506.03981",
    archivePrefix = "arXiv",
    primaryClass = "nucl-ex",
    doi = "10.1140/epjc/s10052-015-3819-5",
    journal = "Eur. Phys. J. C",
    volume = "76",
    number = "3",
    pages = "107",
    year = "2016"
}

@article{Chapon:2020heu,
    author = "Chapon, Emilien and others",
    title = "{Prospects for quarkonium studies at the high-luminosity LHC}",
    eprint = "2012.14161",
    archivePrefix = "arXiv",
    primaryClass = "hep-ph",
    reportNumber = "MIT-CTP/5231, JLAB-THY-20-3240",
    doi = "10.1016/j.ppnp.2021.103906",
    journal = "Prog. Part. Nucl. Phys.",
    volume = "122",
    pages = "103906",
    year = "2022"
}

@article{Kusina:2017gkz,
    author = "Kusina, Aleksander and Lansberg, Jean-Philippe and Schienbein, Ingo and Shao, Hua-Sheng",
    title = "{Gluon Shadowing in Heavy-Flavor Production at the LHC}",
    eprint = "1712.07024",
    archivePrefix = "arXiv",
    primaryClass = "hep-ph",
    doi = "10.1103/PhysRevLett.121.052004",
    journal = "Phys. Rev. Lett.",
    volume = "121",
    number = "5",
    pages = "052004",
    year = "2018"
}

@article{Forte:2010,
    author = "Forte, Stefano and Laenen, Eric and Nason, Paolo and Rojo, Juan",
    title = "{Heavy quarks in deep-inelastic scattering}",
    eprint = "1001.2312",
    archivePrefix = "arXiv",
    primaryClass = "hep-ph",
    doi = "10.1016/j.nuclphysb.2010.03.014",
    journal = "Nucl. Phys. B",
    volume = "834",
    pages = "116--162",
    year = "2010"
}

@article{Harris:1999,
    author = "Harris, B. W. and Laenen, E. and Moch, S. and Smith, J.",
    title = "{Heavy Quark Production in Deep-Inelastic Scattering at HERA}",
    eprint = "hep-ph/9905365",
    archivePrefix = "arXiv",
    primaryClass = "hep-ph",
    month = "5",
    year = "1999"
}

@article{Ablinger:2014vwa,
    author = {Ablinger, J. and Behring, A. and Bl{\"u}mlein, J. and De Freitas, A. and Hasselhuhn, A. and von Manteuffel, A. and Round, M. and Schneider, C. and Wi{\ss}brock, F.},
    title = "{The 3-Loop Non-Singlet Heavy Flavor Contributions and Anomalous Dimensions for the Structure Function $F_2(x,Q^2)$ and Transversity}",
    eprint = "1406.4654",
    archivePrefix = "arXiv",
    primaryClass = "hep-ph",
    reportNumber = "DESY-13--210, DO--TH-13-11, MITP-14-028, SFB-CPP-14-26, LPN-14-074",
    doi = "10.1016/j.nuclphysb.2014.07.010",
    journal = "Nucl. Phys. B",
    volume = "886",
    pages = "733--823",
    year = "2014"
}

@article{H1:2009uwa,
    author = "Aaron, F. D. and others",
    collaboration = "H1",
    title = "{Measurement of the Charm and Beauty Structure Functions using the H1 Vertex Detector at HERA}",
    eprint = "0907.2643",
    archivePrefix = "arXiv",
    primaryClass = "hep-ex",
    reportNumber = "DESY-09-096, DESY09-096",
    doi = "10.1140/epjc/s10052-009-1190-0",
    journal = "Eur. Phys. J. C",
    volume = "65",
    pages = "89--109",
    year = "2010"
}

@article{Abramowicz:2015mha,
  author  = {Abramowicz, H. and others},
  title   = {{Combination of measurements of inclusive deep inelastic \(e^{\pm}p\) scattering cross sections and QCD analysis of HERA data}},
  journal = {Eur. Phys. J. C},
  volume  = {75},
  number  = {12},
  pages   = {580},
  year    = {2015},
  doi     = {10.1140/epjc/s10052-015-3710-4},
  eprint  = {1506.06042},
  archivePrefix = {arXiv},
  primaryClass = {hep-ex},
  collaboration = {H1 and ZEUS}
}

@article{H1:2018flt,
  author  = {H1 and ZEUS Collaborations},
  title   = {{Combination and QCD analysis of beauty and charm production cross-section measurements in deep inelastic \(ep\) scattering at HERA}},
  journal = {Eur. Phys. J. C},
  volume  = {78},
  number  = {6},
  pages   = {473},
  year    = {2018},
  doi     = {10.1140/epjc/s10052-018-5848-3},
  eprint  = {1804.01019},
  archivePrefix = {arXiv},
  primaryClass = {hep-ex}
}

@article{Aaron:2009aa,
  author  = {Aaron, F. D. and others},
  title   = {{Combined Measurement and QCD Analysis of the Inclusive \(e^{\pm}p\) Scattering Cross Sections at HERA}},
  journal = {JHEP},
  volume  = {01},
  pages   = {109},
  year    = {2010},
  doi     = {10.1007/JHEP01(2010)109},
  eprint  = {0911.0884},
  archivePrefix = {arXiv},
  primaryClass = {hep-ex},
  collaboration = {H1 and ZEUS}
}

@article{ZEUS:2014gka,
  author  = {Abramowicz, H. and others},
  title   = {{Measurement of beauty and charm production in deep inelastic scattering at HERA and measurement of the beauty-quark mass}},
  journal = {JHEP},
  volume  = {09},
  pages   = {127},
  year    = {2014},
  doi     = {10.1007/JHEP09(2014)127},
  eprint  = {1405.6915},
  archivePrefix = {arXiv},
  primaryClass = {hep-ex},
  collaboration = {ZEUS}
}

@article{H1:2005vma,
    author = "Aktas, A. and others",
    collaboration = "H1",
    title = "{Measurement of F(2)**c anti-c and F(2)**b anti-b at low Q*2 and x using the H1 vertex detector at HERA}",
    eprint = "hep-ex/0507081",
    archivePrefix = "arXiv",
    reportNumber = "DESY-05-110",
    doi = "10.1140/epjc/s2005-02415-6",
    journal = "Eur. Phys. J. C",
    volume = "45",
    pages = "23--33",
    year = "2006"
}

@article{H1:2004esl,
    author = "Aktas, A. and others",
    collaboration = "H1",
    title = "{Measurement of F2($c \bar{c}$) and F2($b \bar{b}$) at high $Q^{2}$ using the H1 vertex detector at HERA}",
    eprint = "hep-ex/0411046",
    archivePrefix = "arXiv",
    reportNumber = "DESY-04-209",
    doi = "10.1140/epjc/s2005-02154-8",
    journal = "Eur. Phys. J. C",
    volume = "40",
    pages = "349--359",
    year = "2005"
}

@article{ZEUS:2007yva,
    author = "Chekanov, S. and others",
    collaboration = "ZEUS",
    title = "{Measurement of D mesons production in deep inelastic scattering at HERA}",
    eprint = "0704.3562",
    archivePrefix = "arXiv",
    primaryClass = "hep-ex",
    reportNumber = "DESY-07-052",
    doi = "10.1088/1126-6708/2007/07/074",
    journal = "JHEP",
    volume = "07",
    pages = "074",
    year = "2007"
}

@article{H1:2004bwe,
    author = "Aktas, A. and others",
    collaboration = "H1",
    title = "{Inclusive production of D+, D0, D+(s) and D*+ mesons in deep inelastic scattering at HERA}",
    eprint = "hep-ex/0408149",
    archivePrefix = "arXiv",
    reportNumber = "DESY-04-156",
    doi = "10.1140/epjc/s2004-02069-x",
    journal = "Eur. Phys. J. C",
    volume = "38",
    pages = "447--459",
    year = "2005"
}

@article{ZEUS:1997tfb,
    author = "Breitweg, J. and others",
    collaboration = "ZEUS",
    title = "{D* production in deep inelastic scattering at HERA}",
    eprint = "hep-ex/9706009",
    archivePrefix = "arXiv",
    reportNumber = "DESY-97-089, ANL-HEP-PR-98-59",
    doi = "10.1016/S0370-2693(97)00847-2",
    journal = "Phys. Lett. B",
    volume = "407",
    pages = "402--418",
    year = "1997"
}

@article{H1:2015ubc,
    author = "Abramowicz, H. and others",
    collaboration = "H1, ZEUS",
    title = "{Combination of measurements of inclusive deep inelastic ${e^{\pm }p}$ scattering cross sections and QCD analysis of HERA data}",
    eprint = "1506.06042",
    archivePrefix = "arXiv",
    primaryClass = "hep-ex",
    reportNumber = "DESY-15-039",
    doi = "10.1140/epjc/s10052-015-3710-4",
    journal = "Eur. Phys. J. C",
    volume = "75",
    number = "12",
    pages = "580",
    year = "2015"
}

@article{H1:2012xnw,
    author = "Abramowicz, H. and others",
    collaboration = "H1, ZEUS",
    title = "{Combination and QCD Analysis of Charm Production Cross Section Measurements in Deep-Inelastic ep Scattering at HERA}",
    eprint = "1211.1182",
    archivePrefix = "arXiv",
    primaryClass = "hep-ex",
    reportNumber = "DESY-12-172",
    doi = "10.1140/epjc/s10052-013-2311-3",
    journal = "Eur. Phys. J. C",
    volume = "73",
    number = "2",
    pages = "2311",
    year = "2013"
}

@article{Hou:2019efy,
    author = "Hou, Tie-Jiun and others",
    title = "{New CTEQ global analysis of quantum chromodynamics with high-precision data from the LHC}",
    eprint = "1912.10053",
    archivePrefix = "arXiv",
    primaryClass = "hep-ph",
    reportNumber = "MSUHEP-19-025, PITT-PACC-1911, SMU-HEP-19-03",
    doi = "10.1103/PhysRevD.103.014013",
    journal = "Phys. Rev. D",
    volume = "103",
    number = "1",
    pages = "014013",
    year = "2021"
}

@article{Bailey:2020ooq,
    author = "Bailey, S. and Cridge, T. and Harland-Lang, L. A. and Martin, A. D. and Thorne, R. S.",
    title = "{Parton distributions from LHC, HERA, Tevatron and fixed target data: MSHT20 PDFs}",
    eprint = "2012.04684",
    archivePrefix = "arXiv",
    primaryClass = "hep-ph",
    reportNumber = "IPPP/20/58",
    doi = "10.1140/epjc/s10052-021-09057-0",
    journal = "Eur. Phys. J. C",
    volume = "81",
    number = "4",
    pages = "341",
    year = "2021"
}

@article{Laenen:1992zk,
    author = "Laenen, Eric and Riemersma, S. and Smith, J. and van Neerven, W. L.",
    title = "{Complete O (alpha-s) corrections to heavy flavor structure functions in electroproduction}",
    reportNumber = "ITP-SB-92-09",
    doi = "10.1016/0550-3213(93)90201-Y",
    journal = "Nucl. Phys. B",
    volume = "392",
    pages = "162--228",
    year = "1993"
}

@article{H1:2009pze,
    author = "Aaron, F. D. and others",
    collaboration = "H1, ZEUS",
    title = "{Combined Measurement and QCD Analysis of the Inclusive e+- p Scattering Cross Sections at HERA}",
    eprint = "0911.0884",
    archivePrefix = "arXiv",
    primaryClass = "hep-ex",
    reportNumber = "DESY-09-158",
    doi = "10.1007/JHEP01(2010)109",
    journal = "JHEP",
    volume = "01",
    pages = "109",
    year = "2010"
}

@article{Schmitt:2017nwe,
    author = "Schmitt, Stefan",
    editor = "Checchia, Paolo and others",
    collaboration = "H1, ZEUS",
    title = "{Charm and Beauty Production in Deep-inelastic Scattering at HERA}",
    doi = "10.22323/1.314.0400",
    journal = "PoS",
    volume = "EPS-HEP2017",
    pages = "400",
    year = "2017"
}

@article{Eskola:2021nhw,
    author = "Eskola, Kari J. and Paakkinen, Petja and Paukkunen, Hannu and Salgado, Carlos A.",
    title = "{EPPS21: a global QCD analysis of nuclear PDFs}",
    eprint = "2112.12462",
    archivePrefix = "arXiv",
    primaryClass = "hep-ph",
    doi = "10.1140/epjc/s10052-022-10359-0",
    journal = "Eur. Phys. J. C",
    volume = "82",
    number = "5",
    pages = "413",
    year = "2022"
}

@article{Eskola:2009uj,
    author = "Eskola, K. J. and Paukkunen, H. and Salgado, C. A.",
    title = "{EPS09: A New Generation of NLO and LO Nuclear Parton Distribution Functions}",
    eprint = "0902.4154",
    archivePrefix = "arXiv",
    primaryClass = "hep-ph",
    doi = "10.1088/1126-6708/2009/04/065",
    journal = "JHEP",
    volume = "04",
    pages = "065",
    year = "2009"
}

@article{AbdulKhalek:2021gbh,
    author = "Abdul Khalek, R. and others",
    title = "{Science Requirements and Detector Concepts for the Electron-Ion Collider}: {EIC Yellow Report}",
    eprint = "2103.05419",
    archivePrefix = "arXiv",
    primaryClass = "physics.ins-det",
    reportNumber = "BNL-220990-2021-FORE, JLAB-PHY-21-3198, LA-UR-21-20953",
    doi = "10.1016/j.nuclphysa.2022.122447",
    journal = "Nucl. Phys. A",
    volume = "1026",
    pages = "122447",
    year = "2022"
}

@inproceedings{Arleo:2026tpb,
    author = "Arleo, F. and others",
    title = "{2025 EIC-France Workshop: Physics Highlights and Perspectives}",
    booktitle = "{2025 EIC-France Workshop}",
    eprint = "2602.19664",
    archivePrefix = "arXiv",
    primaryClass = "hep-ph",
    month = "2",
    year = "2026"
}

@article{Li:2025lxr,
    author = "Li, Xuan",
    title = "{Recent open heavy flavor studies for the Electron-Ion Collider}",
    eprint = "2501.18044",
    archivePrefix = "arXiv",
    primaryClass = "nucl-ex",
    reportNumber = "LA-UR-24-31401",
    doi = "10.22323/1.465.0113",
    journal = "PoS",
    volume = "QNP2024",
    pages = "113",
    year = "2025"
}

@article{Alwall:2014hca,
    author = "Alwall, J. and Frederix, R. and Frixione, S. and Hirschi, V. and Maltoni, F. and Mattelaer, O. and Shao, H. -S. and Stelzer, T. and Torrielli, P. and Zaro, M.",
    title = "{The automated computation of tree-level and next-to-leading order differential cross sections, and their matching to parton shower simulations}",
    eprint = "1405.0301",
    archivePrefix = "arXiv",
    primaryClass = "hep-ph",
    reportNumber = "CERN-PH-TH-2014-064, CP3-14-18, LPN14-066, MCNET-14-09, ZU-TH-14-14",
    doi = "10.1007/JHEP07(2014)079",
    journal = "JHEP",
    volume = "07",
    pages = "079",
    year = "2014"
}

@article{Malace:2014uea,
    author = "Malace, Simona and Gaskell, David and Higinbotham, Douglas W. and Cloet, Ian",
    title = "{The Challenge of the EMC Effect: existing data and future directions}",
    eprint = "1405.1270",
    archivePrefix = "arXiv",
    primaryClass = "nucl-ex",
    reportNumber = "JLAB-PHY-14-1886",
    doi = "10.1142/S0218301314300136",
    journal = "Int. J. Mod. Phys. E",
    volume = "23",
    number = "08",
    pages = "1430013",
    year = "2014"
}

@article{NewMuon:1993oys,
    author = "Arneodo, M. and others",
    collaboration = "New Muon",
    title = "{A Reevaluation of the Gottfried sum}",
    reportNumber = "CERN-PPE-94-32, CERN-PPE-94-032, CERN-PPE-93-117",
    doi = "10.1103/PhysRevD.50.R1",
    journal = "Phys. Rev. D",
    volume = "50",
    pages = "R1--R3",
    year = "1994"
}

@article{EuropeanMuon:1983wih,
    author = "Aubert, J. J. and others",
    collaboration = "European Muon",
    title = "{The ratio of the nucleon structure functions $F2_n$ for iron and deuterium}",
    reportNumber = "CERN-EP/83-14",
    doi = "10.1016/0370-2693(83)90437-9",
    journal = "Phys. Lett. B",
    volume = "123",
    pages = "275--278",
    year = "1983"
}

@article{SpinMuonSMC:1997mkb,
    author = "Adams, D. and others",
    collaboration = "Spin Muon (SMC)",
    title = "{Spin structure of the proton from polarized inclusive deep inelastic muon - proton scattering}",
    eprint = "hep-ex/9702005",
    archivePrefix = "arXiv",
    reportNumber = "CERN-PPE-97-022, CERN-PPE-97-22, DAPNIA-SPHN-97-27",
    doi = "10.1103/PhysRevD.56.5330",
    journal = "Phys. Rev. D",
    volume = "56",
    pages = "5330--5358",
    year = "1997"
}

@article{HERMES:1998mat,
    author = "Ackerstaff, K. and others",
    collaboration = "HERMES",
    title = "{The HERMES spectrometer}",
    eprint = "hep-ex/9806008",
    archivePrefix = "arXiv",
    reportNumber = "DESY-98-057",
    doi = "10.1016/S0168-9002(98)00769-4",
    journal = "Nucl. Instrum. Meth. A",
    volume = "417",
    pages = "230--265",
    year = "1998"
}

@article{Bjorken:1968dy,
    author = "Bjorken, J. D.",
    title = "{Asymptotic Sum Rules at Infinite Momentum}",
    reportNumber = "SLAC-PUB-0510",
    doi = "10.1103/PhysRev.179.1547",
    journal = "Phys. Rev.",
    volume = "179",
    pages = "1547--1553",
    year = "1969"
}

@book{Devenish:2004pb,
    author = "Devenish, R. and Cooper-Sarkar, A.",
    title = "{Deep inelastic scattering}",
    publisher = "Oxford Univ. Press",
    year = "2004",
    note = " \href{https://doi.org/10.1093/acprof:oso/9780198506713.001.0001}{10.1093/acprof:oso/9780198506713.001.0001}"
}

@article{Gladilin:2014tba,
    author = "Gladilin, Leonid",
    title = "{Fragmentation fractions of $c$ and $b$ quarks into charmed hadrons at LEP}",
    eprint = "1404.3888",
    archivePrefix = "arXiv",
    primaryClass = "hep-ex",
    doi = "10.1140/epjc/s10052-014-3250-3",
    journal = "Eur. Phys. J. C",
    volume = "75",
    number = "1",
    pages = "19",
    year = "2015"
}

@article{Li:2022ewa,
    author = "Li, X. and others",
    title = "{Open Heavy Flavor Studies for the ECCE Detector at the Electron Ion Collider}",
    eprint = "2207.10632",
    archivePrefix = "arXiv",
    primaryClass = "physics.ins-det",
    reportNumber = "LANL report number: LA-UR-22-27181",
    month = "7",
    year = "2022"
}

@article{ParticleDataGroup:2024cfk,
    author = "Navas, S. and others",
    collaboration = "Particle Data Group",
    title = "{Review of particle physics}",
    doi = "10.1103/PhysRevD.110.030001",
    journal = "Phys. Rev. D",
    volume = "110",
    number = "3",
    pages = "030001",
    year = "2024"
}

@article{CMS:2011pdu,
    author = "Chatrchyan, Serguei and others",
    collaboration = "CMS",
    title = "{Measurement of the $B^0$ production cross section in $pp$ Collisions at $\sqrt{s}=7$ TeV}",
    eprint = "1104.2892",
    archivePrefix = "arXiv",
    primaryClass = "hep-ex",
    reportNumber = "CERN-PH-EP-2011-034, CMS-BPH-10-005",
    doi = "10.1103/PhysRevLett.106.252001",
    journal = "Phys. Rev. Lett.",
    volume = "106",
    pages = "252001",
    year = "2011"
}

@article{CMS:2011oft,
    author = "Khachatryan, Vardan and others",
    collaboration = "CMS",
    title = "{Measurement of the $B^+$ Production Cross Section in pp Collisions at $\sqrt{s} = 7${\textasciitilde}TeV}",
    eprint = "1101.0131",
    archivePrefix = "arXiv",
    primaryClass = "hep-ex",
    reportNumber = "CERN-PH-EP-2010-087, CMS-BPH-10-004",
    doi = "10.1103/PhysRevLett.106.112001",
    journal = "Phys. Rev. Lett.",
    volume = "106",
    pages = "112001",
    year = "2011"
}

@article{Kusina:2020dki,
    author = "Kusina, A. and Lansberg, J.-P. and Schienbein, I. and Shao, H.-S.",
    title = "{Reweighted nuclear PDFs using heavy-flavor production data at the LHC}",
    eprint = "2012.11462",
    archivePrefix = "arXiv",
    primaryClass = "hep-ph",
    doi = "10.1103/PhysRevD.104.014010",
    journal = "Phys. Rev. D",
    volume = "104",
    number = "1",
    pages = "014010",
    year = "2021"
}

@article{Duwentaster:2021icx,
    author = "Duwentaster, P. and others",
    title = "{Impact of heavy quark and quarkonium data on nuclear gluon PDFs}",
    eprint = "2105.09872",
    archivePrefix = "arXiv",
    primaryClass = "hep-ph",
    doi = "10.1103/PhysRevD.105.114043",
    journal = "Phys. Rev. D",
    volume = "105",
    number = "11",
    pages = "114043",
    year = "2022"
}

@article{Klasen:2023,
    author = "Klasen, M. and Paukkunen, H.",
    title = "{Nuclear PDFs After the First Decade of LHC Data}",
    journal = "Annu. Rev. Nucl. Part. Sci.",
    volume = "73",
    pages = "321--353",
    year = "2023",
    doi = "10.1146/annurev-nucl-102122-022747"
}

@article{Kovarik:2015cma,
    author = "Kovarik, K. and Kusina, A. and Je{\v{z}}o, T. and Clark, D. B. and Keppel, C. and Lyonnet, F. and Morf{\'i}n, J. G. and Olness, F. I. and Owens, J. F. and Schienbein, I. and Yu, J. Y.",
    title = "{nCTEQ15 -- Global analysis of nuclear parton distributions with uncertainties in the CTEQ framework}",
    eprint = "1509.00792",
    archivePrefix = "arXiv",
    primaryClass = "hep-ph",
    reportNumber = "LPSC-15-153, MS-TP-15-11, FERMILAB-PUB-15-375-ND-PPD-T",
    doi = "10.1103/PhysRevD.93.085037",
    journal = "Phys. Rev. D",
    volume = "93",
    number = "8",
    pages = "085037",
    year = "2016"
}

@article{Muzakka:2022has,
    author = "Muzakka, A. and others",
    collaboration = "nCTEQ",
    title = "{Compatibility of LHC pPb data within the nCTEQ15 framework}",
    eprint = "2204.13157",
    archivePrefix = "arXiv",
    primaryClass = "hep-ph",
    doi = "10.1103/PhysRevD.106.054004",
    journal = "Phys. Rev. D",
    volume = "106",
    number = "5",
    pages = "054004",
    year = "2022"
}

@article{AbdulKhalek:2022fyi,
    author = "Abdul Khalek, R. and Gauld, R. and Giani, T. and Nocera, E. R. and Rabemananjara, T. R. and Rojo, J.",
    title = "{nNNPDF3.0: evidence for a modified partonic structure in heavy nuclei}",
    eprint = "2201.12363",
    archivePrefix = "arXiv",
    primaryClass = "hep-ph",
    journal = "Eur. Phys. J. C",
    volume = "82",
    pages = "507",
    year = "2022"
}

@article{AbdulKhalek:2020yuc,
    author = "Abdul Khalek, Rabah and Ethier, Jacob J. and Rojo, Juan and van Weelden, Gijs",
    title = "{nNNPDF2.0: quark flavor separation in nuclei from LHC data}",
    eprint = "2006.14629",
    archivePrefix = "arXiv",
    primaryClass = "hep-ph",
    reportNumber = "Nikhef/2020-006",
    doi = "10.1007/JHEP09(2020)183",
    journal = "JHEP",
    volume = "09",
    pages = "183",
    year = "2020"
}

@article{AbdulKhalek:2019mzd,
    author = "Abdul Khalek, Rabah and Ethier, Jacob J. and Rojo, Juan",
    collaboration = "NNPDF",
    title = "{Nuclear parton distributions from lepton-nucleus scattering and the impact of an electron-ion collider}",
    eprint = "1904.00018",
    archivePrefix = "arXiv",
    primaryClass = "hep-ph",
    reportNumber = "Nikhef-2019-005",
    doi = "10.1140/epjc/s10052-019-6983-1",
    journal = "Eur. Phys. J. C",
    volume = "79",
    number = "6",
    pages = "471",
    year = "2019"
}

@article{Gerschel:1988wn,
    author = "Gerschel, C. and Hufner, J.",
    title = "{A Contribution to the Description of J / psi Suppression in Nucleus Nucleus Collisions}",
    eprint = "hep-ph/9901416",
    archivePrefix = "arXiv",
    journal = "Phys. Lett. B",
    volume = "207",
    pages = "253--258",
    year = "1988",
    doi = "10.1016/0370-2693(88)90570-9"
}

@article{Vogt:1999cu,
    author = "Vogt, R.",
    title = "{Heavy quark production in nuclear collisions}",
    eprint = "hep-ph/9903551",
    archivePrefix = "arXiv",
    journal = "Prog. Part. Nucl. Phys.",
    volume = "43",
    pages = "197--275",
    year = "1999",
    doi = "10.1016/S0146-6410(99)00096-7"
}

@article{Ferreiro:2014bia,
    author = "Ferreiro, Elena G.",
    title = "{Cold Nuclear Matter Effects on $J/\psi$ production}",
    eprint = "1411.0549",
    archivePrefix = "arXiv",
    primaryClass = "hep-ph",
    journal = "Phys. Lett. B",
    volume = "749",
    pages = "98--103",
    year = "2015",
    doi = "10.1016/j.physletb.2015.07.066"
}

@article{Capella:2005cn,
    author = "Capella, A. and Ferreiro, E. G.",
    title = "{J/psi suppression at RHIC}",
    eprint = "hep-ph/0505032",
    archivePrefix = "arXiv",
    journal = "Eur. Phys. J. C",
    volume = "42",
    pages = "419--424",
    year = "2005",
    doi = "10.1140/epjc/s2005-02302-3"
}

@article{Capella:2000zp,
    author = "Capella, A. and Ferreiro, E. G. and Kaidalov, A. B.",
    title = "{J/psi suppression in p A and A A collisions}",
    eprint = "hep-ph/0002100",
    archivePrefix = "arXiv",
    journal = "Phys. Rev. Lett.",
    volume = "85",
    pages = "2080--2083",
    year = "2000",
    doi = "10.1103/PhysRevLett.85.2080"
}

@article{Gavin:1990gm,
    author = "Gavin, Sean and Vogt, Ramona",
    title = "{J/psi Suppression from Hadronic Collisions to Heavy Ion Collisions}",
    journal = "Nucl. Phys. B",
    volume = "345",
    pages = "104--124",
    year = "1990",
    doi = "10.1016/0550-3213(90)90609-M"
}

@article{Arleo:2012hn,
    author = "Arleo, Francois and Peigne, Stephane",
    title = "{J/$\psi$ suppression in p-A collisions from parton energy loss in cold nuclear matter}",
    eprint = "1204.4609",
    archivePrefix = "arXiv",
    primaryClass = "hep-ph",
    journal = "Phys. Rev. Lett.",
    volume = "109",
    pages = "122301",
    year = "2012",
    doi = "10.1103/PhysRevLett.109.122301"
}

@article{Sharma:2012dy,
    author = "Sharma, Rishi and Vitev, Ivan",
    title = "{High transverse momentum quarkonium production and suppression in heavy ion collisions}",
    eprint = "1203.0329",
    archivePrefix = "arXiv",
    primaryClass = "hep-ph",
    journal = "Phys. Rev. C",
    volume = "87",
    number = "4",
    pages = "044905",
    year = "2013",
    doi = "10.1103/PhysRevC.87.044905"
}

@article{Arleo:2010rb,
    author = "Arleo, Francois and Peigne, Stephane and Sami, Tarek",
    title = "{Revisiting scaling properties of medium-induced gluon radiation}",
    eprint = "1006.0863",
    archivePrefix = "arXiv",
    primaryClass = "hep-ph",
    journal = "Phys. Rev. D",
    volume = "83",
    pages = "114036",
    year = "2011",
    doi = "10.1103/PhysRevD.83.114036"
}

@article{Brodsky:1992nq,
    author = "Brodsky, Stanley J. and Mueller, Alfred H.",
    title = "{Using Nuclei to Probe Short Distance Quark Quark Interactions}",
    journal = "Phys. Lett. B",
    volume = "206",
    pages = "685--690",
    year = "1988",
    doi = "10.1016/0370-2693(88)90719-8"
}

@article{Gavin:1991qk,
    author = "Gavin, Sean and Vogt, Ramona",
    title = "{Energy loss as a source of $J/\psi$ suppression in $p A$ collisions}",
    journal = "Phys. Rev. Lett.",
    volume = "78",
    pages = "1006--1009",
    year = "1997",
    doi = "10.1103/PhysRevLett.78.1006"
}

@article{Brodsky:1989ex,
    author = "Brodsky, Stanley J. and Hoyer, Paul",
    title = "{Nucleus as a Color Filter in QCD}",
    journal = "Phys. Rev. Lett.",
    volume = "63",
    pages = "1566--1569",
    year = "1989",
    doi = "10.1103/PhysRevLett.63.1566"
}

@article{Ducloue:2015gfa,
    author = {Duclou{\'e}, B. and Lappi, T. and M{\"a}ntysaari, H.},
    title = "{Forward $J/\psi$ production in proton-nucleus collisions at high energy}",
    eprint = "1503.02789",
    archivePrefix = "arXiv",
    primaryClass = "hep-ph",
    doi = "10.1103/PhysRevD.91.114005",
    journal = "Phys. Rev. D",
    volume = "91",
    number = "11",
    pages = "114005",
    year = "2015"
}

@article{Ma:2015sia,
    author = "Ma, Y.-Q. and Venugopalan, R. and Zhang, H.-F.",
    title = "{Heavy quarkonium production and polarization in p+A collisions}",
    eprint = "1503.07772",
    archivePrefix = "arXiv",
    primaryClass = "hep-ph",
    journal = "Phys. Rev. D",
    volume = "92",
    pages = "054010",
    year = "2015",
    doi = "10.1103/PhysRevD.92.054010"
}

@article{Fujii:2013gxa,
    author = "Fujii, Hirotsugu and Watanabe, Kazuhiro",
    title = "{Heavy quark pair production in high energy pA collisions: Open heavy flavors}",
    eprint = "1308.1258",
    archivePrefix = "arXiv",
    primaryClass = "hep-ph",
    journal = "Nucl. Phys. A",
    volume = "920",
    pages = "78--93",
    year = "2013",
    doi = "10.1016/j.nuclphysa.2013.10.006"
}

@article{Qiu:2013qka,
    author = "Qiu, Jian-Wei and Sun, Peng and Xiao, Bo-Wen and Yuan, Feng",
    title = "{Universal Cold Nuclear Matter Effects in Quarkonium Production}",
    eprint = "1310.2230",
    archivePrefix = "arXiv",
    primaryClass = "hep-ph",
    journal = "Phys. Rev. D",
    volume = "89",
    number = "3",
    pages = "034007",
    year = "2014",
    doi = "10.1103/PhysRevD.89.034007"
}

@article{Kopeliovich:2001ee,
    author = "Kopeliovich, B. Z. and Tarasov, A. V. and Hufner, J.",
    title = "{Coherence phenomena in charmonium production off nuclei at the energies of RHIC and LHC}",
    eprint = "hep-ph/0104254",
    archivePrefix = "arXiv",
    journal = "Nucl. Phys. A",
    volume = "696",
    pages = "669--714",
    year = "2001",
    doi = "10.1016/S0375-9474(01)01221-7"
}

@article{Ferreiro:2008wc,
    author = "Ferreiro, E. G. and Fleuret, F. and Lansberg, J. P. and Rakotozafindrabe, A.",
    title = "{Cold nuclear matter effects on J/psi production: Intrinsic and extrinsic transverse momentum effects}",
    eprint = "0809.4684",
    archivePrefix = "arXiv",
    primaryClass = "hep-ph",
    journal = "Phys. Lett. B",
    volume = "680",
    pages = "50--55",
    year = "2009",
    doi = "10.1016/j.physletb.2009.08.028"
}

@article{Ferreiro:2011xy,
    author = "Ferreiro, E. G. and Fleuret, F. and Lansberg, J. P. and Rakotozafindrabe, A.",
    title = "{Impact of the nuclear modification of the gluon densities on $J/\psi$ production at RHIC and LHC}",
    eprint = "1111.1228",
    archivePrefix = "arXiv",
    primaryClass = "hep-ph",
    journal = "Phys. Rev. C",
    volume = "88",
    number = "4",
    pages = "047901",
    year = "2013",
    doi = "10.1103/PhysRevC.88.047901"
}

@article{Ferreiro:2013pua,
    author = "Ferreiro, E. G. and Fleuret, F. and Lansberg, J. P. and Rakotozafindrabe, A.",
    title = "{$\Upsilon$ production in $p$Pb collisions at $\sqrt{s_{NN}}=5.02$ TeV}",
    eprint = "1305.4569",
    archivePrefix = "arXiv",
    primaryClass = "hep-ph",
    journal = "Phys. Rev. C",
    volume = "88",
    pages = "047901",
    year = "2013",
    doi = "10.1103/PhysRevC.88.047901"
}

@article{Vogt:2010aa,
    author = "Vogt, R.",
    title = "{Cold Nuclear Matter Effects on Heavy Quarkonium Production}",
    eprint = "1003.3497",
    archivePrefix = "arXiv",
    primaryClass = "hep-ph",
    journal = "Phys. Rev. C",
    volume = "81",
    pages = "044903",
    year = "2010",
    doi = "10.1103/PhysRevC.81.044903"
}

@article{Klasen:2023ugq,
  author        = {Klasen, Michael and Kovarik, Karol and Potthoff, Florian},
  title         = {{Nuclear parton densities and heavy-quark production at the Electron-Ion Collider}},
  journal       = {Phys. Rev. D},
  volume        = {108},
  number        = {1},
  pages         = {014018},
  year          = {2023},
  eprint        = {2305.03548},
  archivePrefix = {arXiv},
  doi           = {10.1103/PhysRevD.108.014018}
}

@article{Segarra:2020gtj,
    author = "Segarra, E. P. and others",
    title = "{Extending nuclear PDF analyses into the high-$x$ , low-$Q^2$ region}",
    eprint = "2012.11566",
    archivePrefix = "arXiv",
    primaryClass = "hep-ph",
    reportNumber = "FERMILAB-PUB-20-668-E, FNAL-PUB-20-668, JLAB-THY-20-3303, SMU-HEP-20-07, MS-TP-20-40, KA-TP-16-2020, MS-TP-20-40,
  KA-TP-16-2020, P3H-20-052, IFJPAN-IV-2020-9",
    doi = "10.1103/PhysRevD.103.114015",
    journal = "Phys. Rev. D",
    volume = "103",
    number = "11",
    pages = "114015",
    year = "2021"
}

@article{Cloet:2019mql,
    author = {Clo{\"e}t, I. C. and others},
    title = "{Exposing Novel Quark and Gluon Effects in Nuclei}",
    eprint = "1902.10572",
    archivePrefix = "arXiv",
    primaryClass = "nucl-ex",
    doi = "10.1088/1361-6471/ab2731",
    journal = "J. Phys. G",
    volume = "46",
    number = "9",
    pages = "093001",
    year = "2019"
}

@article{Arrington:2021vuu,
    author = "Arrington, J. and others",
    title = "{Measurement of the EMC effect in light and heavy nuclei}",
    eprint = "2110.08399",
    archivePrefix = "arXiv",
    primaryClass = "nucl-ex",
    doi = "10.1103/PhysRevC.104.065203",
    journal = "Phys. Rev. C",
    volume = "104",
    number = "6",
    pages = "065203",
    year = "2021"
}

@article{Castro:2026xyr,
    author = "Castro, Santiago and Del Pio, Clara and Kardos, Adam and Moch, Sven-Olaf and Spourdalakis, Aris",
    title = "{Heavy-quark pair-production in DIS at NLO QCD matched to a parton shower}",
    eprint = "2606.21510",
    archivePrefix = "arXiv",
    primaryClass = "hep-ph",
    month = "6",
    year = "2026"
}

@article{Ablinger:2025awb,
    author = {Ablinger, J. and Behring, A. and Bl{\"u}mlein, J. and De Freitas, A. and von Manteuffel, A. and Schneider, C. and Sch{\"o}nwald, K.},
    title = "{The three-loop single-mass heavy-flavor corrections to the structure functions $F_2(x,Q^2)$ and $g_1(x,Q^2)$}",
    eprint = "2509.16124",
    archivePrefix = "arXiv",
    primaryClass = "hep-ph",
    reportNumber = "DESY 25-127, ZU-TH 59/25, MPP-2025-182, RISC Report Series 25-08",
    doi = "10.1016/j.physletb.2026.140540",
    journal = "Phys. Lett. B",
    volume = "878",
    pages = "140540",
    year = "2026"
}

@article{Hekhorn:2024tqm,
    author = "Hekhorn, Felix and Magni, Giacomo and Nocera, Emanuele R. and Rabemananjara, Tanjona R. and Rojo, Juan and Schaus, Adrianne and Stegeman, Roy",
    title = "{Heavy quarks in polarised deep-inelastic scattering at the electron-ion collider}",
    eprint = "2401.10127",
    archivePrefix = "arXiv",
    primaryClass = "hep-ph",
    reportNumber = "Nikhef-2023-010, TIF-UNIMI-2023-27, Edinburgh 2023/2",
    doi = "10.1140/epjc/s10052-024-12524-z",
    journal = "Eur. Phys. J. C",
    volume = "84",
    number = "2",
    pages = "189",
    year = "2024"
}

@article{Anderle:2021hpa,
    author = "Anderle, Daniele Paolo and Dong, Xin and Hekhorn, Felix and Kelsey, Matthew and Radhakrishnan, Sooraj and Sichtermann, Ernst and Xia, Lei and Xing, Hongxi and Yuan, Feng and Zhao, Yuxiang",
    title = "{Probing gluon helicity with heavy flavor at the Electron-Ion Collider}",
    eprint = "2110.04489",
    archivePrefix = "arXiv",
    primaryClass = "hep-ex",
    reportNumber = "PhysRevD.104.114039",
    doi = "10.1103/PhysRevD.104.114039",
    journal = "Phys. Rev. D",
    volume = "104",
    number = "11",
    pages = "114039",
    year = "2021"
}

@phdthesis{Hekhorn:2019nlf,
    author = "Hekhorn, Felix",
    title = "{Next-to-Leading Order QCD Corrections to Heavy-Flavour Production in Neutral Current DIS}",
    eprint = "1910.01536",
    archivePrefix = "arXiv",
    primaryClass = "hep-ph",
    doi = "10.15496/publikation-34811",
    school = "Tubingen U., Math. Inst.",
    year = "2019"
}

@article{Anderle:2023uvi,
    author = "Anderle, Daniele Paolo and Guo, Aiqiang and Hekhorn, Felix and Liang, Yutie and Ma, Yuming and Xia, Lei and Xing, Hongxi and Zhao, Yuxiang",
    title = "{Probing gluon distributions with D0 production at the EicC}",
    eprint = "2307.16135",
    archivePrefix = "arXiv",
    primaryClass = "nucl-ex",
    doi = "10.1103/PhysRevD.109.034021",
    journal = "Phys. Rev. D",
    volume = "109",
    number = "3",
    pages = "034021",
    year = "2024"
}

@article{Buza:1995ie,
    author = "Buza, M. and Matiounine, Y. and Smith, J. and Migneron, R. and van Neerven, W. L.",
    title = "{Heavy quark coefficient functions at asymptotic values Q**2 {\ensuremath{>}}{\ensuremath{>}} m**2}",
    eprint = "hep-ph/9601302",
    archivePrefix = "arXiv",
    reportNumber = "NIKHEF-95-070, ITP-SB-95-59, INLO-PUB-22-95",
    doi = "10.1016/0550-3213(96)00228-3",
    journal = "Nucl. Phys. B",
    volume = "472",
    pages = "611--658",
    year = "1996"
}

@article{Blumlein:2015qcn,
    author = {Bl{\"u}mlein, Johannes},
    title = "{A Kinematic Condition on Intrinsic Charm}",
    eprint = "1511.00229",
    archivePrefix = "arXiv",
    primaryClass = "hep-ph",
    reportNumber = "DESY-15-198, DO-TH-15-15",
    doi = "10.1016/j.physletb.2015.12.068",
    journal = "Phys. Lett. B",
    volume = "753",
    pages = "619--621",
    year = "2016"
}

@article{Brodsky:1980pb,
    author = "Brodsky, S. J. and Hoyer, P. and Peterson, C. and Sakai, N.",
    title = "{The Intrinsic Charm of the Proton}",
    reportNumber = "NORDITA-80-18",
    doi = "10.1016/0370-2693(80)90364-0",
    journal = "Phys. Lett. B",
    volume = "93",
    pages = "451--455",
    year = "1980"
}

@article{Alekhin:2012vu,
    author = {Alekhin, S. and Bl{\"u}mlein, J. and Daum, K. and Lipka, K. and Moch, S.},
    title = "{Precise charm-quark mass from deep-inelastic scattering}",
    eprint = "1212.2355",
    archivePrefix = "arXiv",
    primaryClass = "hep-ph",
    reportNumber = "DESY-12-233, DO-TH-12-35, LPN-12-132, SFB-CPP-12-96",
    doi = "10.1016/j.physletb.2013.02.010",
    journal = "Phys. Lett. B",
    volume = "720",
    pages = "172--176",
    year = "2013"
}

@article{Hou:2017khm,
    author = "Hou, Tie-Jiun and Dulat, Sayipjamal and Gao, Jun and Guzzi, Marco and Huston, Joey and Nadolsky, Pavel and Schmidt, Carl and Winter, Jan and Xie, Keping and Yuan, C. -P.",
    title = "{CT14 Intrinsic Charm Parton Distribution Functions from CTEQ-TEA Global Analysis}",
    eprint = "1707.00657",
    archivePrefix = "arXiv",
    primaryClass = "hep-ph",
    reportNumber = "MSUHEP-17-012",
    doi = "10.1007/JHEP02(2018)059",
    journal = "JHEP",
    volume = "02",
    pages = "059",
    year = "2018"
}

@article{Flore:2020jau,
    author = "Flore, Carlo and Lansberg, Jean-Philippe and Shao, Hua-Sheng and Yedelkina, Yelyzaveta",
    title = "{Large-$P_T$ inclusive photoproduction of $J/\psi$ in electron-proton collisions at HERA and the EIC}",
    eprint = "2009.08264",
    archivePrefix = "arXiv",
    primaryClass = "hep-ph",
    doi = "10.1016/j.physletb.2020.135926",
    journal = "Phys. Lett. B",
    volume = "811",
    pages = "135926",
    year = "2020"
}

\end{document}